\documentclass[prd,aps,tightenlines,aps,showpacs]{revtex4}

\usepackage{amsmath}
\usepackage{verbatim}
\usepackage{graphicx}
\usepackage[usenames]{color}
\usepackage{psfrag}
\usepackage{epstopdf}
\usepackage{hyperref}
\def\beq{\begin{equation}}
\def\eeq{\end{equation}}
\def\bea{\setlength\arraycolsep{1.4pt}\begin{eqnarray}}
\def\eea{\end{eqnarray}}
\def\bit{\begin{itemize}}
\def\eit{\end{itemize}}

\def\tl{\tilde}

\def\h{{\cal H}}

\begin{document}

\title{Anisotropic dark energy and CMB anomalies}
\author{Richard Battye} \email{rbattye@jb.man.ac.uk}

\affiliation{Jodrell Bank Center for Astrophysics, University of Manchester,Manchester, M13 9PL  UK}
\author{Adam Moss} \email{adammoss@phas.ubc.ca}
\affiliation{ Department of Physics \& Astronomy, University of British Columbia,Vancouver, BC, V6T 1Z1  Canada}

\date{\today}

\begin{abstract}
We investigate the breaking of global statistical isotropy caused by a dark energy component with an energy-momentum tensor which has point symmetry, that could represent a cubic or hexagonal crystalline lattice. In such models Gaussian, adiabatic initial conditions created during inflation can lead to anisotropies in the cosmic microwave background whose spherical harmonic coefficients are correlated, contrary to the standard assumption. We develop an adaptation of the line of sight integration method that can be applied to models where the background energy-momentum tensor is isotropic, but whose linearized perturbations are anisotropic. We then show how this can be applied to the cases of cubic and hexagonal symmetry. We compute quantities which show that such models are indistinguishable from isotropic models even in the most extreme parameter choices, in stark contrast to models with anisotropic initial conditions based on inflation. The reason for this is that the dark energy based models contribute to the CMB anistropy via the inegrated Sachs-Wolfe effect, which is only relevent when the dark energy is dominant, that is, on the very largest scales. For inflationary models, however,  the anisotropy is present on all scales.
\end{abstract}

\maketitle

%----------------- INTRODUCTION -----------------------

\section{Introduction} 

Accurate measurements of the anisotropies in the cosmic microwave background (CMB) have lead to a revolution in cosmology. They have provided strong evidence for the basic tenets of the $\Lambda$CDM cosmology. This model has three matter/energy components: baryonic matter, cold dark matter and dark energy in the form of cosmological constant. It also postulates a spectrum of near scale invariant, adiabatic perturbations, thought to be created during inflation. These are usually assumed to been to be compatible with Gaussianity and global statistical isotropy (GSI).

However, both these crucial properties have recently been questioned by detailed analyses of the data from the Wilkinson Microwave Anisotropy Probe (WMAP)~\cite{Hinshaw:2008kr}. Non-Gaussianity is expected at some level if an inflationary epoch is the origin of the adiabatic density perturbations. The present claims of a detection of non-Gaussianity  based on the bispectrum~\cite{Yadav:2007yy} are somewhat higher than predicted by standard slow roll inflation, but could be an indication of non-standard physics during inflation. Violations of GSI, at first sight, appear to be much more worrying, and the analyses which have found such properties have often referred to them as ``CMB anomalies''.

The first such property identified was the low CMB quadropole. This was first noted on the basis of observations from the Cosmic Background Explorer (COBE) satellite~\cite{Smoot:1992td}, but its significance was only seriously questioned when it persisted in the WMAP data. Recent results from the WMAP collaboration claims that it is still compatible with the $\Lambda$CDM model~\cite{Hinshaw:2006ia}. However, these initial claims have spawned a series of paper which have drawn attention to a  number of unusual properties. These features include a near vanishing of the angular correlation function $C(\theta)$ on scales $\theta>60^{\circ}$~\cite{Spergel:2003cb,Copi:2006tu}, asymmetries in the distribution of the power spectrum~\cite{Eriksen:2003db,Hansen:2004vq,Jaffe:2005pw}, extreme cold spots~\cite{Vielva:2003et,Cruz:2004ce} and correlations between the spherical harmonic coeffficients~\cite{de OliveiraCosta:2003pu,Copi:2003kt,Schwarz:2004gk,Prunet:2004zy,Land:2005ad,Land:2006bn}.

Obvious criticisms of these claims are: (1) there is some undentified instrumental or data processing effect which has been ignored in the analysis; (2) the subtraction of the galactic/extragalactic foregrounds has created artifacts in the maps; (3) the statistical significance has been overestimated. However, no such systematic effects have been found and many of the anomalies are seen at lower significance in the COBE data. Moreover they appear independent of frequency and the shear weight of the evidence seems to point to the fact that there is something unusual about the WMAP maps.

It seems, therefore, sensible to consider how the standard $\Lambda$CDM FRW paradigm with initial perturbations created during slow-roll inflation can be modified to provide an explanation for the observed anomalies. A range of ideas based anisotropic inflation have been proposed~\cite{Buniy:2005qm,Gumrukcuoglu:2006xj} which lead to an initial condition based explanation. A specific realization which has generated interest is due to Ackerman, Carroll and Wise (ACW)~\cite{Ackerman:2007nb}, whose model results in a dipole anisotropy in the initial power spectrum which can be parameterized by 
\beq 
P({\bf k}) = P(k) (1+ g_{\star} (\hat{\bf k} \cdot \hat{\bf n})^2)\,,
\eeq
where $\hat{\bf k}$ and $\hat{\bf n}$ are unit vectors in Fourier and real space,  and $g_{\star}$ quantifies the level of anisotropy. The resulting CMB covariance matrix can be written in the form
\beq
C_{\ell_1m_1\ell_2m_2} = \frac{C_{\ell} \delta_{\ell_1 \ell_2} \delta_{m_1 m_2} + g_{\star} \, \xi_{\ell_1m_1\ell_2m_2}   C_{\ell_1 \ell_2}}{1+g_{\star}/3}\,,
\eeq
where the geometric coefficients $\xi_{\ell_1m_1\ell_2m_2}$ couple $\ell_1$ to $\ell_2=\{\ell_1, \ell_1+2 \}$ and $m_1$ to $m_2=\{m_1, m_1+1, m_1+2 \}$. The $C_{\ell, \ell+2}$ can be computed from a modified version of a CMB code such as CMBFAST~\cite{Seljak:1996is} or CAMB~\cite{Lewis:1999bs}. The normalization factor in the denominator ensures that the cylindrical power spectrum ${1\over 2\ell +1}{\sum}_m C_{\ell m\ell m}=C_{\ell}$ is independent of the anisotropy parameter $g_{\star}$. 

A recent analysis~\cite{Groeneboom:2008fz} has used the WMAP V- and W-band maps to constrain the parameter $g_{\star}$. They found statistically significant deviation from $g_\star=0$ with the best fitting value of $g_{\star}=0.15\pm 0.04$ and the preferred axis in the direction of $(l,b)=(110^{\circ},10^{\circ})$. We will use the apparent success of this model in explaining the data  as a benchmark to compare with the alternative models which we will construct. However, we also note the recent work of ref.~\cite{Himmetoglu:2008hx} which claims the ACW model (and more generally any inflationary model driven by a vector field) admits unstable solutions as perturbations cross the horizon.
 
The oberved CMB fluctuations are a convolution of the initial conditions and the transfer function which models the dynamical effects of the expansions of the Universe and the matter/energy components. An obvious alternative to an initial conditions origin for the observed anomalies is that they are due to anistropy in the transfer function. In particular, we explore the possibility that dark energy (with properties different to those of a pure cosmological constant) is responsible. Since dark energy has only made its presence felt since redshifts $z \approx 1$ (if the equation of state $w \approx   -1$, although there is also the possibility of some subdominant early dark energy if the equation of state is dynamical),  it leaves an imprint on the CMB at the large angular scales required. This imprinting occurs via the Integrated Sachs Wolfe (ISW) effect, as photons experience time varying gravitational potentials along the line of sight to the surface of last scattering.

The nature of dark energy is still a mystery (for a review of dark energy models, see ref.~\cite{Copeland:2006wr}). If dark energy is a regular perfect fluid, then the sound speed is equal to $c_{\rm s}=\sqrt{dP/d\rho}=\sqrt{w}$, where the equation of state $w$ is the ratio of pressure $P$ to energy density $\rho$. Since $w<-1/3$ to achieve the observed acceleration, this means that the sound speed would be imaginary, leading to instabilities in the fluid. This work is based on the postulate that dark energy is an {\em elastic} fluid, with non-zero anisotropic stress (models with anisotropic stress have also been investigated in refs.~\cite{Hu:1998kj,Koivisto:2005mm,Mota:2007sz}). The degree of elasticity is controlled by the shear moduli of the fluid. If these moduli are sufficiently large,  the fluid can be stable even when the pressure is negative, making it suitable as a macroscopic model to describe dark energy. For example, a non-zero {\em isotropic} shear modulus $\mu$ modifies the sound speed to $c_{\rm s}^2=w+4\mu/\left[3(1+w)\rho \right]$, and hence if $\mu$ is large enough the sound speed is real. 

It is possible that a number of shear moduli could characterize the elastic properties of the fluid. In this case the dark energy will generate anisotropic perturbations, as the sound speed will be direction dependent (initial studies of dark energy with an anisotropic equation of state $w$ have also been recently carried out~\cite{Koivisto:2007bp,Koivisto:2008ig}). Our work here extends a proof of concept of this idea in ref.~\cite{Battye:2006mb}, in which we computed the evolution of cosmological perturbations. We showed that isotropic initial conditions could lead to anisotropy in the case of cubic symmetry. Here, we calculate the observational effects on large angle CMB fluctuations. 

The paper is organized as follows. In Section~\ref{sec:formalism} we outline the formalism, firstly  reviewing the  perturbation equations required to  compute CMB anisotropies and showing how mode-mixing occurs between scalars, vectors and tensors. We then construct the covariance matrix, which is non-diagonal, as correlations exist between angular modes. We then outline an efficient method to compute this covariance matrix. In Section~\ref{sec:results} we present the numerical results of our computations, in the process comparing these with the ACW model. Finally, we provide a discussion of our results and some concluding remarks.

%----------------- FORMALISM -----------------------

\section{Formalism} \label{sec:formalism}

\subsection{Stress-energy  of an anisotropic medium with isotropic pressure}

The General Relativistic treatment of an elastic medium has been studied in detail by Carter and others~\cite{Carter:1972cq,Carter:1977qf,Carter:1980c,Carter:1982xm}. The primary goal of this work was to understand the dynamics of neutron stars. Recently, we have used a similar approach to study the dynamics of an isotropic elastic medium acting as dark energy in a flat FRW cosmology~\cite{Battye:2007aa}. Here, we consider the anisotropic case in a flat FRW framework, following on from work in ref.~\cite{Battye:2006mb}. The original manifestation of this model was studied in refs.~\cite{Bucher:1998mh,Battye:1999eq}.

The properties of a general elastic medium are characterized by its energy density $\rho$, pressure tensor $P^{\mu \nu}$ and anisotropic stress, which is specified by the shear tensor $\Sigma^{\mu \nu \rho \sigma}$. The first two of these quantities affect the properties of the background space-time {\em and} its perturbations, while the latter affects only the perturbations. The pressure tensor obeys the symmetry and orthogonality relations $P^{\mu \nu}=P^{(\mu \nu)}$ (where the brackets denote symmetrization with respect to the indices),  $P^{\mu \nu} u_{\nu}=0$,  where $u^{\mu} = a^{-1 }(1, 0, 0, 0)$ is the fluid  flow vector, such that there are 6 free components of $P^{\mu \nu}$. The shear tensor obeys the relations $\Sigma^{\mu \nu \rho \sigma} = \Sigma^{(\mu \nu) (\rho \sigma)} = \Sigma^{\rho \sigma \mu \nu}, \, \Sigma^{\mu \nu \rho \sigma}u_{\sigma}=0$, such that 20 components specify the linear shear response to perturbations~\cite{landau:1959}. 

In this work we assume that the pressure tensor is isotropic, so that it is characterized by a single scalar $P$ though $P^{\mu \nu}=P \gamma^{\mu \nu}$, with $P=w \rho$, where the projection tensor is $\gamma^{\mu \nu}=g^{\mu \nu}+u^{\mu} u^{\nu}$ and the flow vector is normalized by $u^{\mu} u_{\mu} =-1$. This condition restricts the free components of  $\Sigma^{\mu \nu \rho \sigma}$, and therefore the class of models which we study. The more general case of an anisotropic pressure tensor would need to be embedded in a Bianchi background, which is beyond the scope of this work.

In the FRW synchronous gauge, the perturbed energy-momentum components of an elastic medium with isotropic pressure are given by~\cite{Battye:2007aa}
\begin{subequations}
\begin{eqnarray}
{\delta T^{0}}_{0} &=&  (\rho + P) \left( \partial_{i} \xi^{i} + h/2 \right)\,, \label{eqn:T00elast} \\
{\delta T^{i}}_{0} &=& -(\rho + P) \dot{\xi}^{i}\,, \label{eqn:Ti0elast}  \\
{\delta T^{i}}_{j} &=& - {\delta^{i}}_{j} \beta  \left( \partial_{k} \xi^{k} + h/2 \right) -  {{{\Sigma^{i}}_{j}}^{k}}_{l}  \left(\partial_{(k} \xi^{l)} +  {h^{l}}_{k}/2 \right)\,, \label{eqn:Tijelast} 
\end{eqnarray}
\end{subequations}
where $h$ is the trace of the metric perturbation and $\beta=(\rho + P) d P / d\rho$ is the bulk modulus of the fluid. The vector $\xi^i$ is the spatial displacement of the fluid worldlines with respect to the background coordinates -- the time part vanishes due to the orthogonality condition $\xi^{\mu} u_{\mu} = 0$. The evolution equation for $\xi^{i}$ is given by the conservation of energy-momentum $\nabla_{\mu} {T^{\mu}}_{\nu}=0$, 
\beq
(\rho + P)( \ddot{\xi}^{i} + \h \dot{\xi}^{i}) - 3 \beta  \h \dot{\xi}^{i} + \partial^j {\delta T^{i}}_{j}=0 \,,
\eeq
where dots denote derivatives with respect to conformal time and $\h$ is the conformal Hubble parameter. The time derivative $\dot{\xi}^{i}$ is the velocity of the fluid.
 
Here we discuss two classes of model with an {\em isotropic} pressure tensor but an {\em anisotropic} shear tensor. For our numerical results we focus on the first of these (elastic dark energy with a cubic  shear tensor).  We emphasize that these models are very different to the multiply-connected spaces which have recently been used in an attempt to explain some of the CMB anomalies (see for example ref.~\cite{Niarchou:2007nn} for recent constraints). Here, we have a flat background with unconnected topology, and an additional  stress-energy component whose perturbations have point symmetry. 

\subsubsection{Cubic symmetry}

\begin{figure}
\centering
%\mbox{\resizebox{0.3\textwidth}{!}{\includegraphics[angle=0]{cube.eps}}}
%\mbox{\resizebox{0.3\textwidth}{!}{\includegraphics[angle=0]{hex_l.eps}}}
\mbox{\resizebox{0.5\textwidth}{!}{\includegraphics[angle=0]{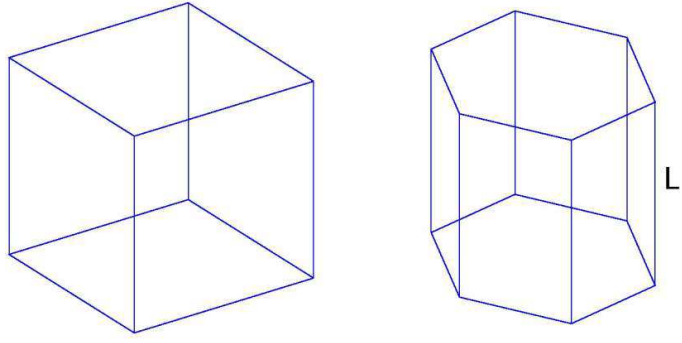}}}
\caption{\label{fig:cell}  Unit cells of two shapes with cubic and hexagonal symmetry.}
\end{figure}

A model with cubic symmetry automatically has an isotropic pressure tensor~\cite{Battye:2005ik} with a single degree of freedom, the pressure scalar, or equivalently the equation of state $w$. There are two degrees of freedom in the shear tensor. If we orient the cubic cell with respect to Cartesian co-ordinates,  the non-zero components are
\begin{subequations} \label{eqn:cube_shear}
\begin{eqnarray}
{{{\Sigma^{x}}_{x}}^{x}}_{x} = {{{\Sigma^{y}}_{y}}^{y}}_{y} = {{{\Sigma^{z}}_{z}}^{z}}_{z} &=& 4\mu_{\rm L}/3\,, \\
{{{\Sigma^{x}}_{x}}^{y}}_{y} = {{{\Sigma^{y}}_{y}}^{z}}_{z} = {{{\Sigma^{z}}_{z}}^{x}}_{x} &=& -2\mu_{\rm L}/3\,, \\
{{{\Sigma^{y}}_{z}}^{y}}_{z} = {{{\Sigma^{x}}_{z}}^{x}}_{z} = {{{\Sigma^{x}}_{y}}^{x}}_{y}& =& \mu_{\rm T}\,,
\end{eqnarray}
\end{subequations}
where $\mu_{\rm L}$ and $\mu_{\rm T}$ are the longitudinal and transverse shear moduli. When $\mu_{\rm L}=\mu_{\rm T}=\mu$ this reduces to the isotropic case. The particular realization of the space-filling cubic unit cell fixes the values of the shear moduli -- the procedure for doing this is outlined in ref.~\cite{Battye:2005ik}, and involves computing the change in energy of the unit cell under spatial transformations. For example, the unit cell of the simple cube shown in the left panel of Fig.~\ref{fig:cell} has $\mu_{\rm L}/\rho=1/6$ and $\mu_{\rm T}/\rho=1/3$. These values apply when the energy density is proportional to the total surface area of the unit cell (a realization being a Nambu-Goto domain wall network with $P=-(2/3)\, \rho$), or when the energy density is proportional to the total edge length of the cell (for example a Nambu-Goto string network with $P=-(1/3) \, \rho$). Values for $\mu_{\rm L}$ and $\mu_{\rm T}$ are computed in ref.~\cite{Battye:2005ik}.

In order to compute the sound speeds at which perturbations propagate, one computes the eigenvalues of the Fresnel tensor, $Q^{\mu \nu}=\Sigma^{\mu \rho \nu \sigma} v_{\rho} v_{\sigma} + \beta v_{\mu} v_{\nu}$, where $v_{\mu}$ is the direction of propagation~\cite{Battye:2005ik}. For the cube, a wave moving parallel to the normal of the faces (for example, $[1,0,0]$) propagates at a speed $v^2= w + 4 \mu_{\rm L}/ \left[ 3 (\rho+P )\right]$, and two other polarization modes propagate in orthogonal directions  at speed $v^2=\mu_{\rm T}/(\rho+P)$. In order to ensure stability and causality, such that $ 0 \le v^2 \le 1$, one finds  a minimum requirement of $\mu_{\rm L}/\rho \ge -3 w(1+w)/4$, and an upper limit of $\mu_{\rm T}/\rho \le 1+ w$. Analysis of waves propagating in other directions gives identical limits with $\mu_{\rm L} \rightarrow \mu_{\rm T}$ and $\mu_{\rm T} \rightarrow \mu_{\rm L}$. The values of $\mu$ which we study will always respect $-3w(1+w)/4={\bar \mu}^{\rm min}\le \mu_{\rm L}/\rho,\mu_{\rm T}/\rho\le {\bar \mu}^{\rm max}=1+w$.

In the cubic case the spatial components of the energy-momentum tensor are then
\begin{equation} \label{eqn:cubic_perturb}
{\delta T^{i}}_{j} = - {\delta^{i}}_{j} ( \beta - 2 \mu_{L}/3 ) \left( \partial_{k} \xi^{k} +  h/2 \right) - 2 \mu_{L} ( \partial_{(j} \xi^{i)} + {h^{i}}_{j}/2) - 2 {S^{i}}_{j}\,, 
\end{equation}
where ${S^{i}}_j$ is the anisotropy source matrix, and is given by
\begin{equation}
{S^{i}}_{j} =   \Delta \mu \left(  \begin{array}{ccc} 0 &  \partial_{(y} \xi^{x)} + {h^{x}}_{y}/2 &  \partial_{(z} \xi^{x)} + {h^{x}}_{z}/2 \\  \partial_{(y} \xi^{x)} + {h^{y}}_{x}/2 & 0  &  \partial_{(y} \xi^{z)} + {h^{y}}_{z}/2 \\  \partial_{(z} \xi^{x)} + {h^{z}}_{x}/2  &  \partial_{(z} \xi^{y)} + {h^{z}}_{y}/2 & 0 \end{array} \right)\,.
\end{equation}
The difference in shear moduli $\Delta \mu = \mu_{\rm T}-\mu_{\rm L} $ quantifies the amount of anisotropy. If $\Delta \mu$ is  non-zero the standard decomposition of perturbations into scalar, vector and tensor (SVT) modes is no longer valid, since ${S^{i}}_j$ cannot be expressed in terms of a single scalar, vector or tensor quantity. This means that mode-mixing occurs and scalar initial conditions can excite vorticity and gravitational waves.

The equation of motion for the fluid displacement vector $\xi^{i}$ is given by
\begin{equation} \label{eqn:cube_eom}
(\rho + P)( \ddot{\xi}^{i} + \h \dot{\xi}^{i}) - 3 \beta  \h \dot{\xi}^{i} - \beta (\partial^{i} \partial_{j} \xi^{j} + \partial^{i} h/2)-\mu_{L} ( \partial^{j} \partial_{j} \xi^{i} + \partial^{i} \partial_{j} \xi^{j}/3 + \partial^{j} {h^{i}}_{j} - \partial^{i} h/3) = V^{i}\,,
\end{equation}
the component $V^{i}$  sourcing anisotropy. This is related to ${S^{i}}_{j}$ via the Einstein equations, and is given by
\begin{equation}
V^{i} =  2 \, \partial^j {S^{i}}_{j} =  \Delta \mu \left(  \begin{array}{ccc} ( \partial_{y} \partial^{y} + \partial_{z} \partial^{z}) \xi^{x} + \partial^{x} (\partial_{y} \xi^{y} + \partial_{z} \xi^{z}) + \partial^{y} {h^{x}}_{y} + \partial^{z} {h^{x}}_{z} \\  ( \partial_{x} \partial^{x} + \partial_{z} \partial^{z}) \xi^{y} + \partial^{y} (\partial_{x} \xi^{x} + \partial_{z} \xi^{z}) + \partial^{x} {h^{y}}_{x} + \partial^{z} {h^{y}}_{z} \\  ( \partial_{x} \partial^{x} + \partial_{y} \partial^{y}) \xi^{z} + \partial^{z} (\partial_{x} \xi^{x} + \partial_{y} \xi^{y}) + \partial^{x} {h^{z}}_{x} + \partial^{y} {h^{z}}_{y} \end{array} \right)\,.
\end{equation}

\subsubsection{Hexagonal symmetry}

Unlike a cubic system, the pressure tensor is not automatically isotropic in the hexagonal case. There are three degrees of freedom in the shear tensor. If we align the $C_6$ symmetry axis with respect to the $z$ co-ordinate,  the non-zero components are
\begin{subequations} \label{eqn:hex_shear}
\begin{eqnarray}
{{{\Sigma^{x}}_{x}}^{x}}_{x} = {{{\Sigma^{y}}_{y}}^{y}}_{y}  = 4\mu /3\,, \quad {{{\Sigma^{z}}_{z}}^{z}}_{z} &=& 4\mu_{\rm A}/3\,, \\
 {{{\Sigma^{x}}_{x}}^{y}}_{y} = -2\mu/3\,, \quad {{{\Sigma^{y}}_{y}}^{z}}_{z} = {{{\Sigma^{z}}_{z}}^{x}}_{x} &=& -2\mu_{\rm A}/3\,, \\
 {{{\Sigma^{x}}_{y}}^{x}}_{y} = \mu \,, \quad {{{\Sigma^{y}}_{z}}^{y}}_{z} = {{{\Sigma^{x}}_{z}}^{x}}_{z} &= &\mu_{\rm B} \,.
\end{eqnarray}
\end{subequations}
In the $x-y$ plane the pressure and shear tensor are isotropic. This is a property of 2D hexagonal structures~\cite{Battye:2005ik}. 

In order to satisfy isotropy of the pressure tensor in 3D, the properties of the space-filling unit cell are not completely free. The unit cell of a simple hexagonal prism is shown in the right panel of Fig.~\ref{fig:cell}. If the edges surrounding the top and bottom hexagonal faces have unit length, the length $L$ must be 
fixed so the pressure tensor is isotropic. If the energy density is proportional to the total edge length (such that $P=-(1/3) \,\rho$), $L$ must also be of unit length, and if the energy density is proportional to the total surface area (such that $P=-(2/3) \, \rho$) then $L=2 \sqrt{3}$. In both cases one finds $\mu/\rho=1/4$, $\mu_{\rm A}/\rho=3/16$ and $\mu_{\rm B}/\rho=1/3$.

As in the cubic case one can compute the eigenvalues of the Fresnel tensor to derive constraints on the values of $\mu$. A mode moving in the direction $[1,0,0]$ propagates at $v^2= w + 4 \mu/ \left[ 3 (\rho+P )\right]$, with two other polarization states propagating in directions $[0,1,0]$ and $[0,0,1]$ at speeds $v^2=\mu/(\rho+P)$ and $v^2=\mu_{\rm B}/(\rho+P)$  respectively. Similarly, a wave in the $[0,0,1]$ direction has a speed $v^2= w + 4 \mu_{\rm A}/ \left[ 3 (\rho+P )\right]$, with two polarization states  with 
$v^2=\mu_{\rm B}/(\rho+P)$. There is no fixed direction to derive lower limits on $\mu_{\rm B}$ and upper limits on $\mu_{\rm A}$, so one should solve the eigenvalue solutions to ensures the solutions respect stability and causality. These conditions are satisfied for the isotropic hexagonal cell discussed above.

In the hexagonal case the spatial components of the energy momentum tensor are given by~(\ref{eqn:cubic_perturb}), with $\mu_{\rm L} \rightarrow \mu$,  and an anisotropic source term
 \begin{equation}
{S^{i}}_{j} =   \Delta \mu_{\rm B}  \left(  \begin{array}{ccc} 0 &  0 &   \partial_{(z} \xi^{x)} + {h^{x}}_{z}/2 \\  0 & 0  &   \partial_{(y} \xi^{z)} + {h^{y}}_{z}/2 \\    \partial_{(z} \xi^{x)} + {h^{z}}_{x}/2  &   \partial_{(z} \xi^{y)} + {h^{z}}_{y}/2 & 0 \end{array} \right) -  \frac{\Delta \mu_{\rm A}}{3}  \left[ {\delta^i}_j \left( \partial_z \xi^z + h_{zz}/2\right)  +  {\delta^{z}}_j{\delta_{z}}^i  \left( \partial_k \xi^k + h/2\right) \right] ,
\end{equation}
where $\Delta \mu_{\rm A} = \mu_{\rm A}-\mu$ and $\Delta \mu_{\rm B} = \mu_{\rm B}-\mu$. 
 
\subsection{Scalar-Vector-Tensor decomposition}

We now require a convenient basis to expand the perturbed Einstein and fluid equations. Scalar, vector and tensor (SVT) quantities can be expanded in terms of an orthogonal set of basis functions $e^{\alpha}_i$ ($\alpha=1 \ldots 3$) in Fourier space. We set $e^1_i=\hat{k}_i$, so that the fluid displacement vector is given by (see for example refs.~\cite{Lifshitz:1963ps,Hu:1997mn,Pereira:2007yy} for further details of this decomposition)
\beq
\xi_{i} = \xi^{\rm (0)} \hat{k}_{i} + \xi^{ (-1)} e_{i}^2 + \xi^{ (+1)} e_{i}^3\,,
\eeq
with analogous expressions for other vector quantities. Here $(0)$ denotes the scalar component and $(\pm 1)$ the two vector modes. The metric perturbation in Fourier space is given by
\begin{equation} 
h_{ij} = \hat{k}_{i} \hat{k}_{j} h + \left( \hat{k}_{i} \hat{k}_{j} - \frac{1}{3} \delta_{ij} \right) 6 \eta + 2 \hat{k}_{(i} h_{j)}^{V} + h_{ij}^{T}\,,
\end{equation} 
with $\hat{k}^{i} h^{V}_{i} = \hat{k}^{i} h_{ij}^{T} = h^{T}_{ii} = 0$. The vector and tensor components of the metric perturbation are then constructed from the basis vectors by
\begin{subequations}
\begin{eqnarray}
h_{i}^{V} &=&h^{ (-1)} e^{2}_{i}  + h^{ (+1)} e^{3}_{i}\,, \\
h_{ij}^{T} &=& h^{(-2)} (e^{2}_{i} e^{2}_{j} - e^{3}_{i} e^{3}_{j}) + h^{(+2)} (e^{2}_{i}e^{3}_{j}+e^{3}_{i}e^{2}_{j})\,,
\end{eqnarray}
\end{subequations}
with $(\pm 2)$ denoting the two tensor modes. For computational purposes we will use the particular choice of $e^{2}_i$ and $e^{3}_i$ (which satisfy the orthogonality condition) given by
\begin{equation} \label{eqn:basis}
e^{2}_i= \frac{1}{\sqrt{\hat{k}_{x}^{2} +\hat{k}_{y}^{2}}} \left(  \begin{array}{ccc} &\hat{k}_{y}&  \\  - &\hat{k}_{x}&  \\  &0& \end{array} \right) ,\hspace{0.1cm}
e^{3}_i= \frac{1}{\sqrt{\hat{k}_{x}^{2} +\hat{k}_{y}^{2}}}  \left(  \begin{array}{ccc} &\hat{k}_{x} \hat{k}_{z}&  \\ &\hat{k}_{y} \hat{k}_{z}&  \\    -&(\hat{k}_{x}^{2} +\hat{k}_{y}^{2})&  \end{array} \right)\,.
\end{equation}

The fluid equations of motion can now be obtained by taking the scalar product of~(\ref{eqn:cube_eom}) with the basis vectors. This is equivalent to decomposing SVT quantities in terms of eigenfunctions of the Laplacian as in refs.~\cite{Lifshitz:1963ps,Hu:1997mn}. However, in this decomposition SVT modes do not decouple due to the anisotropic source term. For the cubic case they are given by (with analogous expressions for the hexagonal case)
\begin{subequations}
\begin{eqnarray}
(\rho + P)( \ddot{\tilde \xi}^{\rm (0)} + \h \dot{\tilde\xi}^{\rm (0)}) - 3 \beta  \h \dot{\tilde\xi}^{\rm (0)} + \left( \beta + 4 \mu_{L}/3 \right) \left( k^{2} {\tilde\xi}^{\rm (0)} +  k h/2 \right) + 4 \mu_{L} k \eta &=&  e^{1}_{i} {\tilde V}^{i}\,, \\
(\rho + P)( \ddot{\tilde\xi}^{\rm (-1)} + \h \dot{\tilde\xi}^{\rm (-1)}) - 3 \beta  \h \dot{\tilde\xi}^{\rm (-1)} + \mu_{L} \left( k^{2} {\tilde\xi}^{\rm (-1)} +  k h^{\rm  (-1)} \right) &=&  e^{2}_{i} {\tilde V}^{i}\,, \\
(\rho + P)( \ddot{\tilde\xi}^{\rm (+1)} + \h \dot{\tilde\xi}^{\rm (+1)}) - 3 \beta  \h \dot{\tilde\xi}^{\rm (+1)} + \mu_{L} \left( k^{2} {\tilde\xi}^{\rm (+1)} +  k h^{\rm (+1)} \right) &=&  e^{3}_{i} {\tilde V}^{i}\,,
\end{eqnarray}
\end{subequations}
where ${\tilde\xi}=i \xi$. The source vector ${\tilde V}^{i}$ for the cubic case is
\begin{equation}
{\tl V}^{i} =   \Delta \mu \left(  \begin{array}{ccc} -( k_{y}^2 + k_{z}^2) {\tl \xi}^{x}  -  k_{x} (k_{y} {\tl \xi}^{y} + k_{z} {\tl \xi}^{z}) - k_{y} h_{xy} - k_{z} {h_{xz}} \\  -( k_x^2 + k_z^2) {\tl \xi}^{y} - k_{y} (k_{x} {\tl \xi}^{x} + k_{z} {\tl \xi}^{z}) - k_{x} {h_{yx}} - k_z {h_{yz}} \\  -( k_x^2 + k_y^2) {\tl \xi}^{z} - k_{z} (k_{x} {\tl \xi}^{x} + k_{y} {\tl \xi}^{y}) - k_{x} {h_{zx}} - k_{y} {h_{zy}} \end{array} \right)\,.
\end{equation}
In the synchronous gauge the Einstein equations are given by
\begin{subequations}
\begin{eqnarray}
a^{2} {G^{0}}_{0} &=&   - 3 {\cal H}^2 - {\cal H} \dot{h} + \partial_{i} \partial^{i} h/2 - \partial_{i} \partial_{j} h^{ij}/2 \,, \label{eqn:G00} \\
2 a^{2} {G^{0}}_{i}&=&   \partial_{i}\dot{h}  - \partial_{j}{\dot{h}^{j}}_{\,\,i} \,, \label{eqn:G0i} \\
2 a^{2} {G^{i}}_{0}&=&   \partial_{j}\dot{h}^{ij} - \partial^{i}\dot{h} \,, \label{eqn:Gi0} \\
a^{2} {G^{i}}_{j}&=&    \left( 2 \dot{{\cal H}} - {\cal H}^{2} \right) {\delta^{i}}_{j} + \left( {\ddot{h}^{i}}_{\,\,j} - \ddot{h} {\delta^{i}}_{j} \right)/2  \label{eqn:Gij}  + {\cal H} \left({\dot{h}^{i}}_{\,\,j} - \dot{h} {\delta^{i}}_{j} \right) + \left({\delta^{i}}_{j} \partial_{k} \partial^{k} h  - \partial_{k} \partial^{k} {h^{i}}_{j} \right) /2   \\ \nonumber
 & & +  \delta^{ik}/2 \left( \partial_{k} \partial_{l} {h^{l}}_{j} + \partial_{j} \partial_{l} {h^{l}}_{k} - \partial_{k} \partial_{j} h  \right)  -  {\delta^{i}}_{j} \partial_{k} \partial_{l} h^{kl} /2 \,. 
\end{eqnarray}
\end{subequations}
Similarly, projecting the Einstein equations with the basis vectors gives the set of first order (constraint) and second order (evolution) equations in Fourier space:

\medskip
\noindent
Constraint:
\begin{subequations} \label{eqn:constraint}
\begin{eqnarray}
 \h \dot{h}/2 - k^{2} \eta  &=& 4 \pi G a^{2} \delta \rho\,, \\
k \dot{\eta} &=&  4 \pi G a^{2} ( \rho + P ) \dot{\tl \xi}^{\rm (0)}\,, \\
k \dot{h}^{\rm (\pm 1)} &=&  16 \pi G a^{2} ( \rho + P ) \dot{\tl \xi}^{\rm (\pm 1)}\,, 
\end{eqnarray}
\end{subequations}
Evolution:
\begin{subequations} \label{eqn:evolution}
\begin{eqnarray}
\ddot{h} + 2 \h \dot{h} - 2k^{2} \eta &=& - 24 \pi G a^{2} \delta P\,, \\
\ddot{h} + 6 \ddot{\eta} + 2 \h ( \dot{h} + 6 \dot{\eta}) - 2 k^{2} \eta &=& - 16 \pi G a^{2}  \Pi^{\rm (0)}\,,\\
\ddot{h}^{\rm (\pm 1)}+2 \h \dot{h}^{\rm (\pm 1)} &=&  - 8 \pi G a^{2}  \Pi^{\rm (\pm 1)}\,, \\
 \ddot{h}^{(\pm 2)} + 2 \h \dot{h}^{(\pm 2)} + k^{2} h^{(\pm 2)}&=& 8 \pi G a^{2}  \Pi^{(\pm 2)}\,, 
 \end{eqnarray}
\end{subequations}
where the sources are
\begin{subequations}  \label{eqn:sources}
\begin{eqnarray}
\delta \rho  &=& - (\rho +P) \left( k {\tl \xi}^{\rm (0)} +  h/2 \right), \\
 \delta P &=& {dP\over d \rho}\delta\rho\,, \\
\Pi^{\rm (0)} &=& 2 \mu_{L} \left( k {\tl \xi}^{(0)} +  h/2 + 3 \eta \right) + 3   e^{1}_{i} e^{j}_{1}  S^{i}_{j}\,,\\
\Pi^{\rm (-1)} &=& 2  \mu_{L} \left(k {\tl \xi}^{\rm (-1)} + h^{\rm (-1)} \right) +4   e^{1}_{i} e^{j}_{2} {S^{i}}_{j} \,,\\
\Pi^{\rm (+1)} &=& 2  \mu_{L} \left(k {\tl \xi}^{\rm (+1)} + h^{\rm (+1)} \right) +4   e^{1}_{i} e^{j}_{3} {S^{i}}_{j} \,,\\
\Pi^{(-2)} &=&  -2  \mu_{L}  h^{(-2)} - 2  \left[ e^{2}_{i} e^{j}_{2} -   e^{3}_{i} e^{j}_{3} \right] {S^{i}}_{j}\,,\\
\Pi^{(+2)} &=& -2  \mu_{L}  h^{(+2)} -2 \left[  e^{2}_{i} e^{j}_{3} + e^{3}_{i} e^{j}_{2}  \right] {S^{i}}_{j}\,.
\end{eqnarray}
\end{subequations}

We have modified the CAMB software~\cite{Lewis:1999bs} to include an anisotropic elastic component in an otherwise standard cosmology. This involves treating each of the $(\pm)$ sources separately, and evolving SVT modes simultaneously due to the coupling (normally SVT modes decouple so can be computed independently). Furthermore, since the source term ${S^{i}}_j$ is dependent on the {\em direction} as well as the magnitude of the ${\bf k}$ vector in Fourier space, we evolve equations at a discrete set of directions and magnitudes of ${\bf k}$. We give more details of this procedure in Section~\ref{sec:approx}.

%----------------- SERIES SOLUTIONS -----------------------

\subsection{Angular dependence induced by anisotropy} \label{sec:ang_k}

\begin{figure}
\centering
%\mbox{\resizebox{0.8\textwidth}{!}{\includegraphics[angle=0]{cube_set.eps}}}
\mbox{\resizebox{0.8\textwidth}{!}{\includegraphics[angle=0]{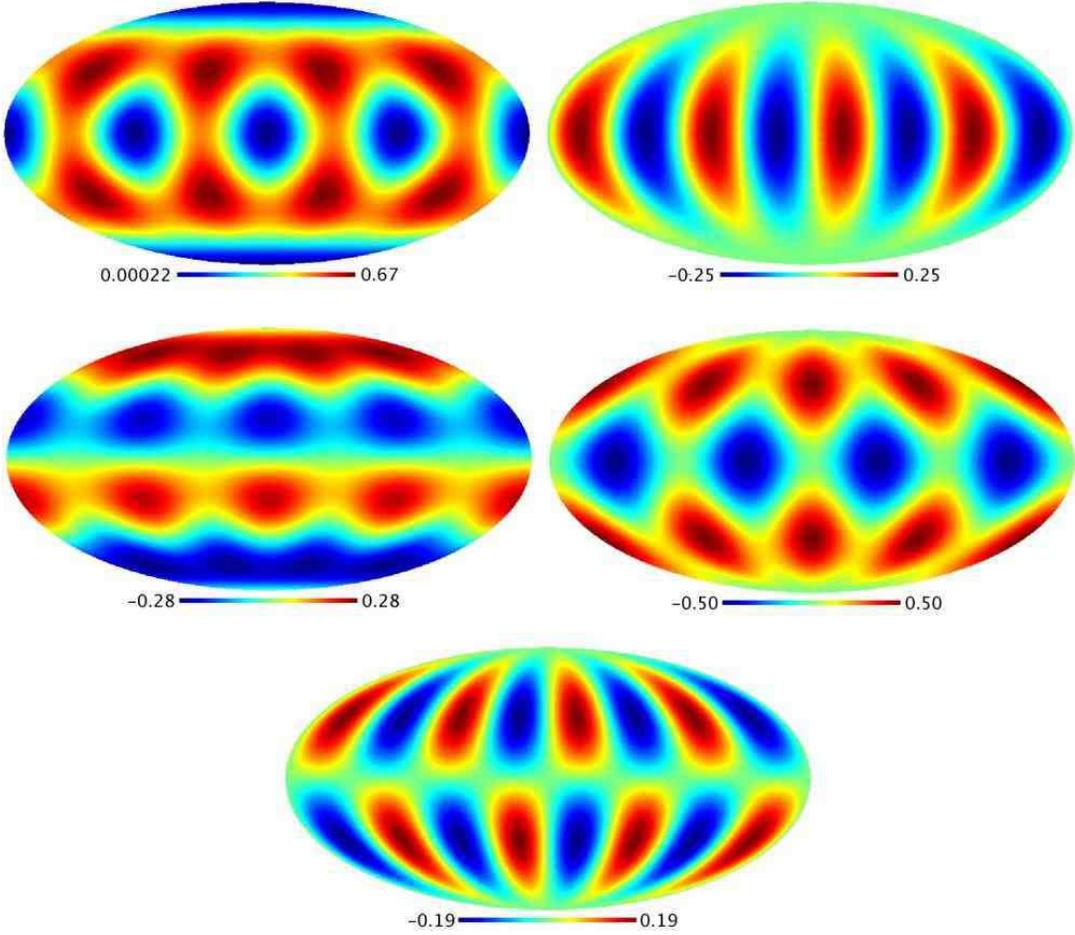}}}
\caption{\label{fig:cub_sources}  Angular dependence in $k$-space for the $(0,-1,+1,-2,+2)$ cubic source functions (top-left, top-right, midddle-left, middle right and bottom respectively), with $\Delta \mu$ arbitrarily set to unity. We have used the equal area Aitoff projection.}
\end{figure}

It is easy to see the angular dependence induced by the anisotropy by considering the projection of the source term ${S^{i}}_j$ with the basis vectors. For an initial scalar curvature fluctuation in the metric, $h_{ij}= (\hat{k}_{i} \hat{k}_{j} - \frac{1}{3} \delta_{ij} ) \, 6 \eta$, the cubic and hexagonal source terms have the form
\begin{equation}
S_{ij}^{\rm init, \, cub} \propto \Delta \mu  \left(  \begin{array}{ccc} 0 & \hat{k}_{x} \hat{k}_{y}  &  \hat{k}_{x} \hat{k}_{z}  \\  \hat{k}_{x} \hat{k}_{y}  & 0  &  \hat{k}_{y} \hat{k}_{z}  \\  \hat{k}_{x} \hat{k}_{z}   &  \hat{k}_{y} \hat{k}_{z}  & 0 \end{array} \right)\,, \quad S_{ij}^{\rm init, \, hex} \propto \Delta \mu_{\rm B}  \left(  \begin{array}{ccc} 0 & 0  &  \hat{k}_{x} \hat{k}_{z}  \\ 0 & 0  &  \hat{k}_{y} \hat{k}_{z}  \\  \hat{k}_{x} \hat{k}_{z}   &  \hat{k}_{y} \hat{k}_{z}  & 0 \end{array} \right) -  \frac{\Delta \mu_{\rm A}}{3}  \delta_{iz}\delta_{jz} \left[ \hat{k}_z^2 - \frac{1}{3} \right] \,.
\end{equation}
Angular mode functions can then be defined by projecting the source term
\begin{subequations}  \label{eqn:proj_sources}
\begin{eqnarray}
S^{\rm (0)} (\theta_{\hat k},\phi_{\hat k}) &=& e^{1}_{i} e^{1}_{j}  S^{\rm init}_{i j} \,, \\
S^{\rm (-1)} (\theta_{\hat k},\phi_{\hat k}) = e^{1}_{i} e^{2}_{j}  S^{\rm init}_{i j}\,, \quad S^{\rm (+1)} (\theta_{\hat k},\phi_{\hat k}) &=& e^{1}_{i} e^{3}_{j}  S^{\rm init}_{i j}\,, \\
S^{(-2)} (\theta_{\hat k},\phi_{\hat k}) = (e^{2}_{i} e^{2}_{j} - e^{3}_{i} e^{3}_{j} )S^{\rm init}_{i j}\,, \quad S^{(+2)} (\theta_{\hat k},\phi_{\hat k}) &=& (e^{2}_{i} e^{3}_{j} + e^{3}_{i} e^{2}_{j}) S^{\rm init}_{i j}\,. 
\end{eqnarray}
\end{subequations}
These cubic source functions are displayed in an Aitoff projection in Fig.~\ref{fig:cub_sources}. It is interesting to note the morphology of the maps in each case. The $(0, -1,-2)$ modes have the full set of cubic symmetries, the scalar term having obvious symmetry by  inspecting the terms for $S^{\rm (0)} \propto  (\hat{k}_{x} \hat{k}_{y} )^2 +(\hat{k}_{x} \hat{k}_{z} )^2+(\hat{k}_{y} \hat{k}_{z} )^2  $. However, the mirror symmetry along the $k_z$-axis is broken for the $(+1,+2)$ modes.  No vector or tensor fluctuations are generated towards the centre of a cubic face or corner - these correspond to the direction of the maximal and minimal scalar fluctuations.

\begin{figure}
\centering
%\mbox{\resizebox{0.8\textwidth}{!}{\includegraphics[angle=0]{hex_set.eps}}}
\mbox{\resizebox{0.8\textwidth}{!}{\includegraphics[angle=0]{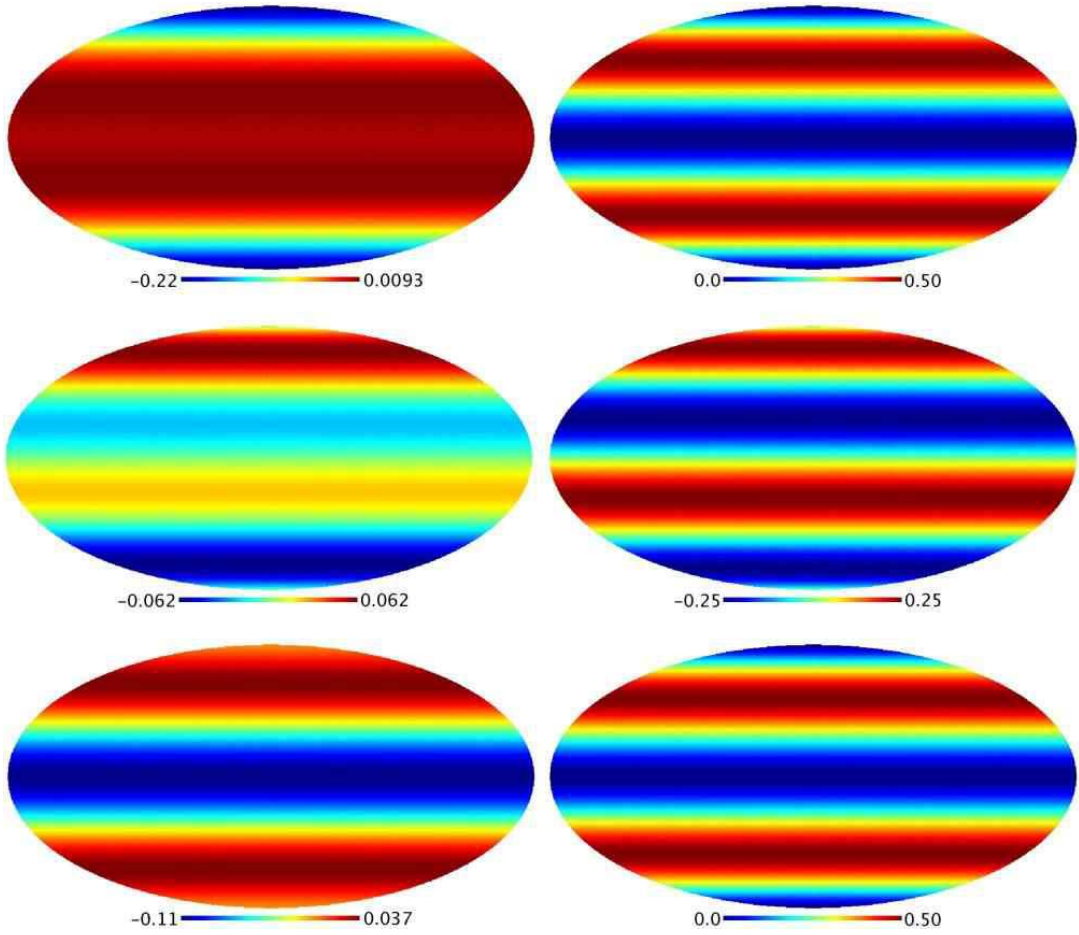}}}
\caption{\label{fig:hex_sources}  Angular dependence in $k$-space for the $(0,+1,-2)$ hexagonal source functions (top, middle and bottom), with $\Delta \mu_{\rm A}=1$ and   $\Delta \mu_{\rm B}=0$ (left set of three), and $\Delta \mu_{\rm A}=0$ and   $\Delta \mu_{\rm B}=1$ (right set of three).}
\end{figure}

The hexagonal source has two terms of different weight depending on $\Delta \mu_{\rm A}$ and $\Delta \mu_{\rm B}$. In all cases though, $S^{\rm (-1)} =S^{(+2)}=0$. The non-zero sources are shown in Fig.~\ref{fig:hex_sources} and as expected, are isotropic in the $k_x-k_y$ plane. 

In order to see the effect of the anisotropic sources on the perturbation equations, we can write down a series expansion in conformal time $\tau$ of the coupled differential equations. For concreteness, we consider a universe consisting of cold dark matter (c), photons ($r$) and an elastic fluid component (e) with $w=-2/3$ and a cubic shear tensor. We assume the photon moments with $\ell > 1$ are zero, which is a good approximation until tight coupling breaks down. In the expansion terms below we suppress the CDM component for clarity (the hexagonal case has $\mu_{\rm L} \rightarrow \mu$):
\begin{subequations}  \label{eqn:series}
\begin{eqnarray}
{\tl \xi}^{\rm (0)}_{(e)} &=& -\frac{k^3 \tau^4}{144} \left( 4 \bar{\mu}_{\rm L} - 1 + 12  S^{\rm (0)}\right)  \nonumber \\
&+& \left \{ \frac{1}{432} \left( \frac{1}{36} + \bar{\mu}_{\rm L}^2 - \frac{11}{30} \bar{\mu}_{\rm L} \right)  + \frac{ S^{\rm (0)}}{72} \left(  \bar{\mu}_{\rm L}  - \frac{11}{60}\right) + \frac{ 1}{192} \left( 4S^{\rm (0), 2} + S^{\rm (-1), 2} + S^{\rm (+1), 2}\right) \right \} k^5 \tau^6 \,, \\
{\tl \xi}^{\rm (\pm 1)}_{(e)}  &=& -\frac{k^3 \tau^4}{24}  S^{\rm (\pm 1)}\,, \\
h &=&  \frac{1}{2} (k \tau)^2 - \frac{1}{216} (k \tau)^4 \,, \\
\eta &=& 1 - \frac{1}{36} (k \tau)^2 + \frac{1}{3240} (k \tau)^4 - \frac{1}{5040}  \omega_{\rm e} \sqrt{\omega_{\rm r}}  k^2 \tau^5 \left [56 (\bar{\mu}_{\rm L}+ 3 S^{\rm (0)} ) -31 \right ]\,, \\
h^{\rm (\pm 1, \pm 2)} &=& -\frac{1}{15}   \omega_{\rm e} \sqrt{\omega_{\rm r}}  k^2 \tau^5 S^{\rm (\pm 1, \pm 2)} \,, 
\end{eqnarray}
\end{subequations}
where $\bar{X}=X / \rho$ and  $\omega_{\rm x}= \Omega_{\rm x} H_{0}^{2}$. One can see how anisotropy in the fluid induces anisotropy in the metric fluctuations, which in turn are a source of large-angle CMB temperature fluctuations via the ISW effect. Additionally, vector and tensor metric components are excited due to mode-mixing.

%----------------- COVARIANCE MATRIX -----------------------

\subsection{CMB covariance matrix}

The CMB temperature is given by $ T({\bf{{x}}},{{\hat{\bf{n}}}},\eta) = T(\eta)\left[1+\Delta({\bf{{x}}},{{\hat{\bf{n}}}},\eta) \right]$, where the perturbation is expanded in terms of spherical harmonics $Y_{\ell m }({{\hat{\bf{n}}}})$,
\beq
\Delta({\bf{{x}}},{{\hat{\bf{n}}}},\eta)=\sum_{\ell=0}^{\infty} \sum_{m=-\ell}^{\ell} a_{\ell m} ({\bf{{x}}},\eta) Y_{\ell m }({{\hat{\bf{n}}}}) \,.
\eeq
Fourier transforming this expression gives the multipole coefficients
\beq
a_{\ell m}({\bf x},\tau_0)={1\over (2\pi)^3}\int d^3{\bf k} \, e^{i{\bf k}\cdot{\bf x}}\int d\Omega_{{{\hat{\bf{n}}}}} \,Y_{\ell m}^*(\hat{\bf n}) \Delta({\bf k},{{\hat {\bf n}}},\tau)\,.
\eeq
The line of sight integral solution for the photon perturbations is
\beq
\Delta({\bf k},\hat{\bf n},\tau_0)=\sum_p\zeta_p({\bf k})\sum_n F^{(n)}(\hat{\bf k},\hat{\bf n})\int_0^{\tau_0}d\tau e^{ix\mu}T^{(n,p)}({\bf k},\tau)\,,
\eeq
where $x=k(\tau_0-\tau)$ and $\mu={\hat{\bf k}}\cdot{\hat{\bf n}}$. The label $p$ (between -2 and 2) represents the different varieties of initial conditions with 0 corresponding to scalars, $\pm 1$ the two vector modes and $\pm 2$ the two tensor modes. The source function $T^{(n,p)}({\bf k},\tau)$ is that computed as a function of position in $k$-space (not just $k=|{\bf k}|$). It represents the response of the $n$-th mode to initial condition type $p$ - all five modes will generally be excited by any particular initial condition. $\zeta_p$ is a random variable describing the properties of type-$p$ initial conditions.

The functions $F^{(n)}$ represent properties of  the scalar, vector and tensor modes and are given by 
\begin{subequations}
\begin{eqnarray}
F^{(0)}(\hat{\bf k},\hat{\bf n})&=& 1\,, \\
F^{(-1)}(\hat{\bf k},\hat{\bf n})&=&{e}^2_i n_i \,,\\
F^{(+1)}(\hat{\bf k},\hat{\bf n})&=&{e}^3_i n_i \,,\\
F^{(-2)}({\hat{\bf k}},\hat{{\bf n}})&=&({e}^2_i {e}^2_j-{e}^3_i {e}^3_j)\hat{n}_i\hat{n}_j\,, \\
F^{(+2)}(\hat{\bf k},\hat{\bf n})&=&({e}^2_i { e}^3_j+{e}^2_j {e}^3_i)\hat{n}_i\hat{n}_j\,.
\end{eqnarray}
\end{subequations}
Hence, we can deduce that 
\beq 
a_{\ell m}({\bf x},\tau_0)={1\over (2\pi)^3}\sum_{n,p}\int d^3{\bf k}\, e^{i{\bf k}\cdot{\bf x}}\zeta_{p}({\bf k})\int_0^{\tau_0}d\tau T^{(n,p)}({\bf k},\tau)I^{(n)}_{\ell m}(x,{\hat {\bf{k}}})\,,
\eeq
where 
\beq 
I_{\ell m}^{(n)}(x,{\hat {\bf{k}}})=\int d\Omega_{\hat{n}}Y_{\ell m}^*({\hat{\bf n}})e^{ix\hat{\bf k}\cdot\hat{\bf n}}F^{(n)}({\hat{\bf k}},{\hat{\bf n}})\,.
\eeq
The $I_{\ell m}^{(n)} (x,{\hat {\bf{k}}})$ functions can be written in terms of linear combinations $J_{\ell m}^{(n)}$ of the spin harmonics (as described in appendix~\ref{app:cov}) by
\beq 
I_{\ell m}^{(n)} (x,{\hat {\bf{k}}}) = 4\pi i^{\ell-|n|}{j_{\ell}(x)\over x^{|n|}} J_{\ell m}^{(n)}(\hat{\bf k})\,.
\eeq

We will assume that the initial conditions are Gaussian random initial conditions and therefore
\beq
\langle \zeta_{p_1}^*({\bf k}_1)\zeta_{p_2}({\bf k}_2)\rangle=(2\pi^2)^2P_{p_1}(k_1)\delta^{(3)}({\bf k}_1-{\bf k}_2)\delta_{p_1p_2}\,.
\label{ps}
\eeq
One can now compute the correlation matrix
\beq \label{eqn:covmat}
C_{\ell_1m_1\ell_2m_2} =\langle a_{\ell_1 m_1}^*({\bf x},\tau_0)a_{\ell_2 m_2}({\bf x},\tau_0)\rangle 
=\sum_{p,n_1,n_2}\int k^2dk P_{p}(k) \Delta_{\ell_1m_1\ell_2m_2}^{n_1n_2p}(k)\,,
\eeq
where 
\beq \label{eqn:cmat}
\Delta_{\ell_1m_1\ell_2m_2}^{n_1n_2p}(k)=(-i)^{\ell_1-|n_1|}i^{\ell_2-|n_2|}\int d\Omega_{\hat k} J_{\ell_1m_1}^{(n_1)*}(\hat {\bf{k}})J_{\ell_2m_2}^{(n_2)}(\hat {\bf{k}})\Delta_{\ell_1}^{(n_1,p)}({\bf k})\Delta_{\ell_2}^{(n_2,p)}({\bf k})\,,
\eeq
and 
\beq 
\Delta^{(n,p)}_{\ell}({\bf k})=\int_0^{\tau_0}d\tau T^{(n,p)}({\bf k},\tau){j_{\ell}[k(\tau_0-\tau)]\over [k(\tau_0-\tau)]^{|n|}}\,.
\eeq
In order to represent the covariance matrix, it will be convenient to replace $(\ell, m)$  with a single index $s=\ell(\ell+1)+m$, that is $C_{s_1 s_2}=C_{\ell_1m_1\ell_2m_2}$ for $s_1=\ell_1(\ell_1+1)+m_1$ and $s_2=\ell_2(\ell_2+1)+m_2$. 

%----------------- APPROXIMATE TREATMENT-----------------------

\subsection{Approximate treatment} \label{sec:approx}

The linearized calculation of section~\ref{sec:ang_k} suggests that for a small anisotropy coefficent (that is, small $\Delta \mu$'s) we can write the transfer functions as
\beq \label{eqn:tran_decomp}
\Delta_{\ell}^{(n, p)}({\bf k})= \delta_{n\, p} A_{\ell}^{(p)} (k) +  B^{(n,p)}_{\ell} (k) S^{(n)}  (\theta_{\hat k},\phi_{\hat k})\,,
\eeq
where $B_{\ell}^{(-1,p)}=B_{\ell}^{(+1,p)}$ and $B_{\ell}^{(-2,p)}=B_{\ell}^{(+2,p)}$. This consists of an isotropic piece $ A_{\ell}^{(p)} (k)$, which is non-zero only for the initial condition type $p$, and an anisotropic piece $ B^{(n,p)}_{\ell} (k) $ which is non-zero for all modes. This approximation significantly reduces computational time, as only several evaluations of $\Delta_{\ell}^{(n, p)}({\bf k})$ are required at each position in $k$-space for each $k=|{\bf k}|$. With knowledge of the source function $S^{(n)}(\theta_{\hat k},\phi_{\hat k})$, this is sufficient to extract the amplitude of the isotropic and anisotropic components of the transfer function. From this, one can evaluate the integral~(\ref{eqn:cmat}) to compute the correlation matrix elements.

To save further computational time, each time the covariance matrix is evaluated (for example, with a different set of cosmological parameters), the integrals over the anisotropic source functions will be the same. Therefore, one can pre-compute matrices of the form 
\beq \label{eqn:ang}
\Pi_{s s^{\prime}}^{n_1n_2}=(-i)^{\ell_1-|n_1|}i^{\ell_2-|n_2|}\int d\Omega_{\hat k} J_{\ell_1m_1}^{(n_1)*}J_{\ell_2m_2}^{(n_2)} S^{(n_1)} S^{(n_2)} \,.
\eeq
With some simple algebra the correlation matrix~(\ref{eqn:covmat}) can be expanded in terms of  these geometric matrix elements, and the angular information~(\ref{eqn:ang}) can be supplemented with amplitudes to give the contributions to the correlation matrix. For the case of cubic symmetry, these elements couple $\ell_1$ to $\ell_2=\{\ell_1, \ell_1+2, \ell_1+4 \}$ and $m_1$ to $m_2=\{m_1, m_1+1, m_1+4, m_1+8 \}$. The resulting matrix is sparse, and we make use of this for storage purposes. A naive approach would lead to the storage of $\mathcal{O} (\ell_{\rm max}^4)$ elements, most of them zero, while the more efficient method only scales as  $\mathcal{O} (\ell_{\rm max}^2)$.

\begin{figure}
\centering
%\mbox{\resizebox{0.99\textwidth}{!}{\includegraphics[angle=0]{04_028_018_fit.eps}}}
\mbox{\resizebox{0.99\textwidth}{!}{\includegraphics[angle=0]{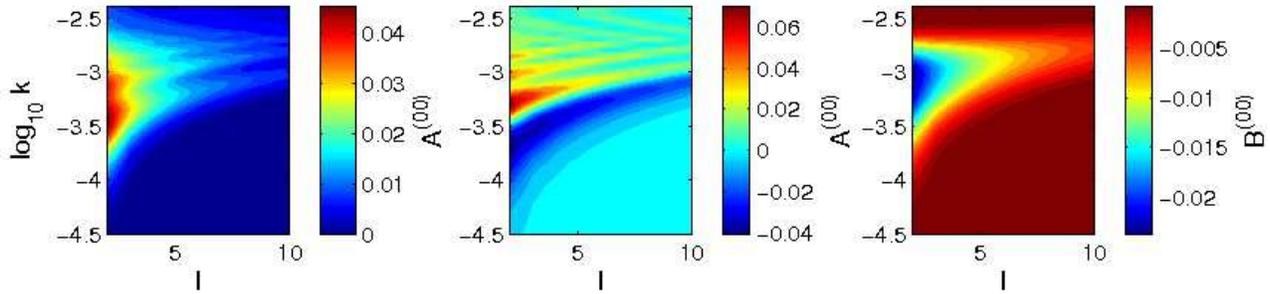}}}
\caption{\label{fig:fit}  Isotropic contribution to the scalar transfer function $\Delta_{\ell}({\bf k})$ for the ISW component (left) and total ISW + last-scattering surface (LSS) components (middle). Also shown is the {\em anisotropic}  component (right) which is generated only via the ISW effect. We use a model with $w=-0.4$, ${\bar\mu}_{\rm T}={\bar\mu}^{\rm min}$ and ${\bar\mu}_{\rm L}={\bar \mu}^{\rm min}+0.1$, where ${\bar\mu}^{\rm min}=0.18$.  }
\end{figure}

In doing this, one also gains some insight into the relative contribution between the isotropic and anisotropic components. In Fig.~\ref{fig:fit} we show the  both the ISW and total ISW + last scattering surface (LSS) components of the scalar transfer function, for a model with $w=-0.4$, ${\bar\mu}_{\rm T}=\mu_{\rm T}/\rho={\bar\mu}^{\rm min}$ and ${\bar \mu}_{\rm L}=\mu_{\rm L}/\rho={\bar\mu}^{\rm min}+0.1$, where ${\bar\mu}^{\rm min}=0.18$.  The justification for using these values is discussed in the following section, but using $| \Delta {\bar \mu} | = 0.1$ allows one to use the approximate treatment outlined here rather than the full numerical evolution. The covariance matrix to  $\ell_{\rm max} \sim 40$ can be computed in only several minutes with this method, compared to the full evolution, which takes $\sim 1$ day. We find errors of less than 5$\%$ for any matrix element given the various $w$ and $| \Delta {\bar \mu}|= 0.1$ which we consider when compared to the full treatment that is significantly more time consuming.

%----------------- RESULTS -----------------------

\section{Results} \label{sec:results}

We now compute the covariance matrix for anisotropic elastic dark energy (AEDE) models with a cubic shear tensor. Our baseline cosmological parameters are those listed for the best fitting $\Lambda$CDM model in ref.~\cite{Dunkley:2008ie}, with  elastic dark energy taking the place of a cosmological constant. We use a scalar comoving curvature perturbation as our initial conditions and we investigate models with $w=-0.2$,  $-0.4$, $-0.6$ and $-0.8$. When we vary the value of $w$, we also adjust the value of the Hubble parameter, $H_0 = 100 \, h\, {\rm km \, sec^{-1} \, Mpc^{-1}}$,  to keep the CMB peak positions consistent with the $\Lambda$CDM model. We also keep the physical matter and baryon densities $\Omega_{\rm m} h^2$ and $\Omega_{\rm b} h^2$  constant, which ensures the CMB peak height also remain fixed. In this way the only difference in the CMB correlation matrix between models occurs at low $\ell$. 

Dark energy effects the ISW contribution to CMB anisotropies in two ways. The first is by its effect on the expansion rate (and the induced decay of gravitational potentials as the total equation of state changes). The second is by the growth of imhomogenities in the dark energy itself~\cite{Weller:2003hw,Bean:2003fb}, in which the sound speed plays an important role.  The case of an isotropic elastic dark energy was discussed in ref.~\cite{Battye:2007aa}. The situation here is similar, except we now have a anisotropic sound speed along with a source of large-scale vector and tensor perturbations.

We find that the largest deviations from isotropy occur for the case $w=-0.2$. This is due to two reasons -- as $w$ approaches $-1$ dark energy perturbations are suppressed, and a higher value of $w$ results in dark energy domination at earlier times. The anisotropic contribution to the transfer function then extends to smaller scales, and hence contributes to higher $\ell$ in the covariance matrix. This value of $w$ would be incompatible with recent SN observations (see for example ref.~\cite{Kowalski:2008ez}), but serves as a baseline for the size of deviations from isotropy that can be generated in these models. We tried using even more extreme values of $w>-0.2$, but in this case the size of the effect actually {\em decreased}. The reason for this is there is some overlap in the $k$ range of the ISW and LSS components of the transfer function, which can be seen by examining Fig.~\ref{fig:fit}. This overlap results in partial cancellation of the isotropic-anisotropic cross term, which is strongest for cases with $w>-0.2$.

For the shear moduli, the minimum and maximum values of $\bar\mu$ which we considered were ${\bar \mu}^{\rm min}=-3w(1+w)/4$, ${\bar \mu}^{\rm max}=1+w$, due to stability and causality constraints. We obtained roughly the same results by interchanging ${\bar \mu}_{\rm T}$ and ${\bar \mu}_{\rm L}$, and found  maximal deviation from isotropy by fixing either one to its minimum value. For the other value, it was  convenient to use $| \Delta{\bar \mu} |=0.1$ for several reasons: (1)  This allowed the approximate treatment to be used, which was computationally much faster and sufficiently accurate; (2) $| \Delta {\bar \mu} |= 0.1$ corresponds to roughly the level of anisotropy found in domain wall lattices, which could be a possible realization of anisotropic dark energy~\cite{Battye:2005ik}; (3) the level of anisotropy does not increase significantly for $| \Delta {\bar \mu} | > 0.1$.
 
The reason for the last point can be understood by considering the effect of the dark energy sound speed on the scalar ISW source term. In the {\em isotropic case}, the difference between a model with $c_{\rm s}^2=0$ and $0.1$ is much larger than, say between $c_{\rm s}^2=0.1$ and $1$~\cite{Battye:2007aa}. In the anisotropic case, fixing one of the  $\mu$'s to zero is equivalent to an anisotropic scalar sound speed which is zero in several spatial directions, and a maximal value of $c_{\rm s}^2=4 |\Delta {\bar \mu} | /[ 3 (1+w)] $ in other directions, with an appropriate interpolation between the two. The scalar source term is therefore rather insensitive to $| \Delta {\bar \mu} | > 0.1 $. A caveat here is that vector and tensor modes are sourced by the anisotropy, whose size are proportional to $\Delta {\bar \mu}$. However, we find that the vector and tensor contributions to the covariance matrix decrease as a function of $\ell$ faster than the anisotropic scalar contribution, and the scalar terms dominate for all but the  lowest $\ell$ modes. 

In the following discussion then, we will set ${\bar\mu}_{\rm T}={\bar\mu}^{\rm min}$  and ${\bar\mu}_{\rm L}={\bar\mu}^{\rm min}+0.1$. We will frequently compare the AEDE results with those of the ACW model. For this purpose, we will orient the preferred axis of the ACW dipole anisotropy along the equator, with $\theta_{\star}=0$ and $\phi_{\star}=90 ^{\circ}$ (see ref.~\cite{Ackerman:2007nb} for notation). In this case, the $\{m, m+1\}$ correlations of the covariance matrix vanish.

\begin{figure}
\centering
%\mbox{\resizebox{0.49\textwidth}{!}{\includegraphics[angle=0]{02_022_012_isw_clmat2.eps}}}
%\mbox{\resizebox{0.49\textwidth}{!}{\includegraphics[angle=0]{02_022_012_full_clmat2.eps}}}
%\mbox{\resizebox{0.49\textwidth}{!}{\includegraphics[angle=0]{ACW_g1_clmat2.eps}}}
%\mbox{\resizebox{0.49\textwidth}{!}{\includegraphics[angle=0]{ACW_g01_clmat2.eps}}}
\mbox{\resizebox{0.9\textwidth}{!}{\includegraphics[angle=0]{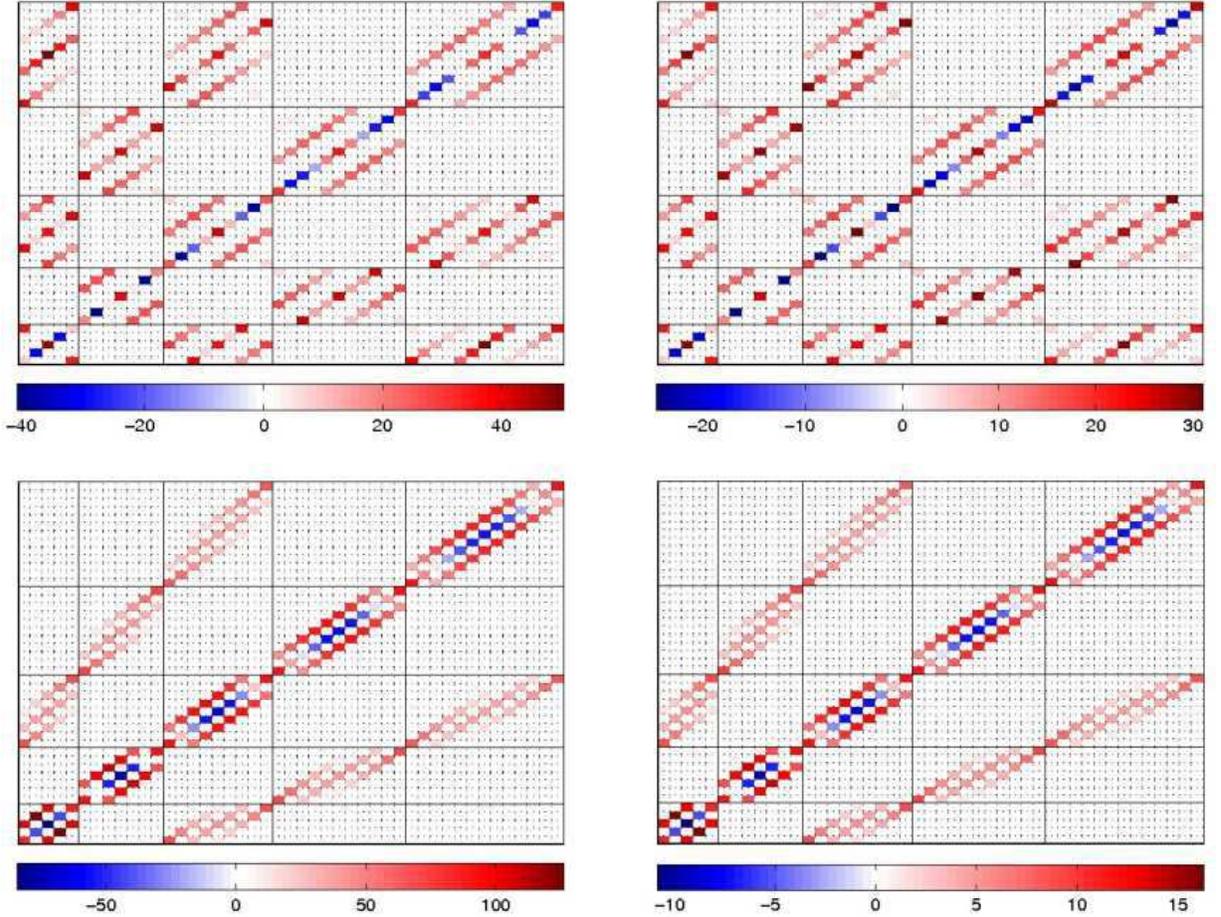}}}
\caption{\label{fig:cov_modes} Representation of the CMB correlation matrix (see text for details) in $(\mu K)^2$ for  the ISW component (top-left) and the total ISW + LSS (top-right), for an AEDE  model with $w=-0.2$, ${\bar\mu}_{\rm T}={\bar\mu}^{\rm min}$  and ${\bar\mu}_{\rm L}={\bar\mu}^{\rm min}+0.1$. Also shown is the ACW correlation matrix for $g_{\star}=1$ (bottom-left) and $g_{\star}=0.1$ (bottom-right).}
\end{figure}

\begin{figure}
\centering
\mbox{\resizebox{0.5\textwidth}{!}{\includegraphics[angle=0]{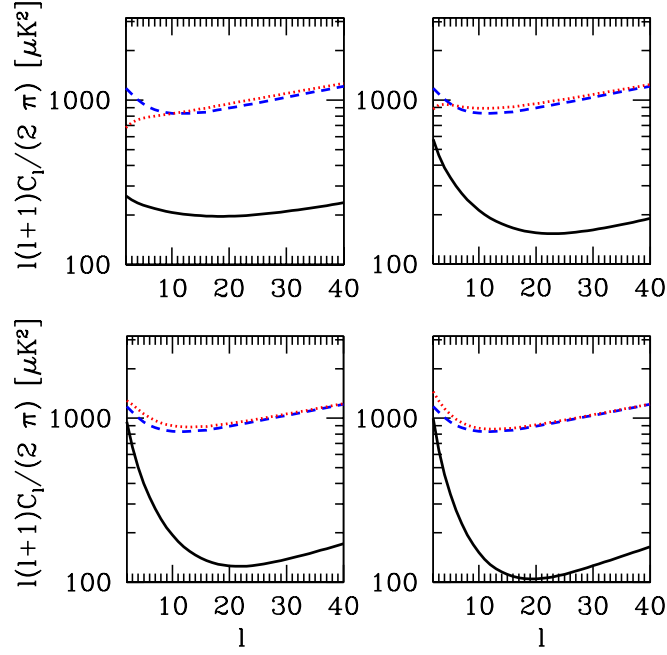}}}
\caption{\label{fig:tt} Cylindrical power spectrum $C_{\ell}$ for $w=-0.2$ (top-left), $w=-0.4$ (top-right), $w=-0.6$ (bottom-left) and $w=-0.8$ (bottom-right).  The solid curve shows the ISW contribution for ${\bar\mu}_{\rm T}={\bar\mu}^{\rm min}$  and ${\bar\mu}_{\rm L}={\bar\mu}^{\rm min}+0.1$ (the exception is for $w=-0.8$, where $\Delta {\bar\mu}=0.08$ due to causality and stability bounds on the sound speed), and the dotted curve shows the total ISW+LSS component. Also shown for comparison is the cylindrical  $C_{\ell}$ for the ACW model (dashed curve).}
\end{figure}

The resulting numerical computation of the total CMB covariance matrix for $\ell=2 \ldots 6$ is shown in Fig.~\ref{fig:cov_modes} for both the ACW and an AEDE model with $w=-0.2$. For visual clarity, we plot the expression
\beq \label{eqn:cov_plot}
\left( |C_{s_1 s_2}| - \delta_{s_1 s_2} C_{s_1 s_1}^{\rm ISO} \right) \sqrt{\ell_1 \ell_2 \left(\ell_1+1 \right) \left(\ell_2+1 \right)}/ (2\pi)\,.
\eeq
Notice that we use the norm of the covariance matrix, since this matrix can (and does in the AEDE case) have imaginary components, resulting from the $i$ prefactor in eqn.~(\ref{eqn:cmat}). This is perfectly natural and this poses no problem for creating map realizations, since the matrix is Hermitian. We subtract the isotropic contribution of the covariance matrix in order to show the anisotropic components more clearly. The isotropic part can be computed by calculating the cylindrical $C_{\ell}$, defined by
\beq \label{eqn:cl}
C_{\ell}=\frac{1}{2\ell+1} \sum_{m} C_{\ell m \ell m}\,.
\eeq
These $C_\ell$'s are shown for a range of models in Fig.~\ref{fig:tt}, which we discuss in more detail below. The remaining $\ell$ factors in eqn.~(\ref{eqn:cov_plot}) are the analogous quantity to the usual $\ell (\ell+1)/(2 \pi)$ when plotting the CMB power spectrum. 

Referring to Fig.~\ref{fig:cov_modes}, the thicker lines inside each plot correspond to the $(2\ell+1)$ blocks for each $m$ mode. We show the ISW and total ISW+LSS components for the AEDE model, and also compare these to the ACW covariance matrix with $g_{\star} =1$ and $0.1$. There are several features to notice about these plots.  Mode couplings occur in the ACW model between $\{ \ell,\ell+2\}$ and in the AEDE model between $\{ \ell,\ell+2\}$ and $\{ \ell,\ell+4\}$. These couplings are expected due to cubic symmetry. The planar angular dependence of anisotropy for the ACW model is also apparent, due to our choice of preferred axis. This results in more power in $m=\pm \ell$ modes  for each $\ell$, which is apparent along the main diagonal of the matrix.

The size of the anisotropic terms from $\ell=2 \ldots 6$ are comparable between the ISW only AEDE model with $w=-0.4$ and ACW model with $g_{\star} =1$. Furthermore, these terms are larger when considering the full ISW+LSS spectrum compared to the ACW model with $g_{\star}=0.1$. However, one also notices a much faster decrease in the size of the terms as a function of $\ell$ compared to ACW, since anisotropy is only sourced at low $\ell$ in the AEDE model.

Another important feature to notice is the change in amplitude of the anisotropic terms in the AEDE model when considering the full ISW+LSS spectrum. 
 These are reduced by a factor of $\sim 2$ compared to the ISW only case, so one {\em cannot} simply add an isotropic $C_{\ell}$ component - the cross term between the ISW+LSS is also important, as we discussed previously. The reduction in amplitude results from a partial cancellation in the isotropic-anisotropic cross term when compared to the ISW only case.
 
Fig.~\ref{fig:tt} shows the cylindrical $C_{\ell}$'s for a range of AEDE models with $w=-0.2$,  $-0.4$, $-0.6$ and $-0.8$, again using ${\bar\mu}_{\rm T}={\bar\mu}^{\rm min}$  and ${\bar\mu}_{\rm L}={\bar\mu}^{\rm min}+0.1$ (except for $w=-0.8$ where we use $\Delta{\bar\mu}=0.08$ to comply with the causality constraint). This quantity is approximately the same for models with switched longitudinal and transverse shear moduli, which changes the sign of the anisotropy parameter $\Delta {\bar\mu}$. The effect of switching the sign of $\Delta {\bar\mu}$ on the covariance matrix results in the $\{ \ell,\ell+2\}$ and $\{ \ell,\ell+4\}$ couplings remaining roughly constant, while the diagonal terms change sign (whilst keeping the $C_{\ell}$ approximately constant).

\begin{figure}
\centering
\mbox{\resizebox{0.4\textwidth}{!}{\includegraphics[angle=0]{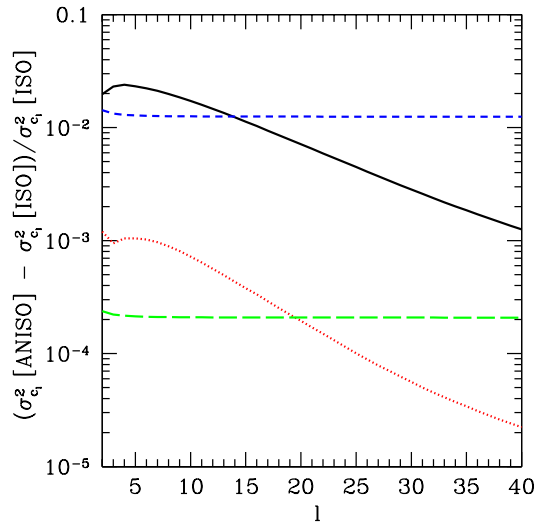}}}
\caption{\label{fig:variance} Fractional increase in cosmic variance over an isotropic model for (solid/dotted) the ISW/total contribution to the power spectrum,  with $w=-0.2$, ${\bar\mu}_{\rm T}={\bar\mu}^{\rm min}$ and ${\bar\mu}_{\rm L}={\bar\mu}^{\rm min}+0.1$. Also shown is ACW result with $g_{\star}=1$ (short-dashed) and $g_{\star}=0.1$ (long-dashed).}
\end{figure}

In anisotropic models it is known that the cosmic variance of the $C_{\ell}$ is greater than an isotropic model with the same $C_{\ell}$~\cite{Ferreira:1997wd}. Physically, this is due a smaller number of degrees of freedom at each $\ell$ mode, that is the distribution is no longer $\chi^2_{2\ell +1}$. The variance of the $C_{\ell}$ can be found by diagonalizing the dimension $(2\ell+1)$ square matrix for each $\ell$, which contains information on the $m$ correlations. The result of this diagonalization, $ \tilde{C}_{\ell m \ell m}$, can then be used to compute the variance using
\beq 
\sigma^2_{C_{\ell}}=\frac{2}{(2\ell+1)^2} \sum_{m} \tilde{C}_{\ell m \ell m}^2\,.
\eeq
In Fig.~\ref{fig:variance} this is compared to the expected variance in an isotropic theory, $\sigma^{2 \,\, \rm ISO}_{C_{\ell}}=2 C_{\ell}^2/(2\ell+1)$. It is again apparent  how the AEDE model approaches the isotropic limit much faster than the ACW model, but neither is particularly significant.

\begin{figure}
\centering
\mbox{\resizebox{0.49\textwidth}{!}{\includegraphics[angle=0]{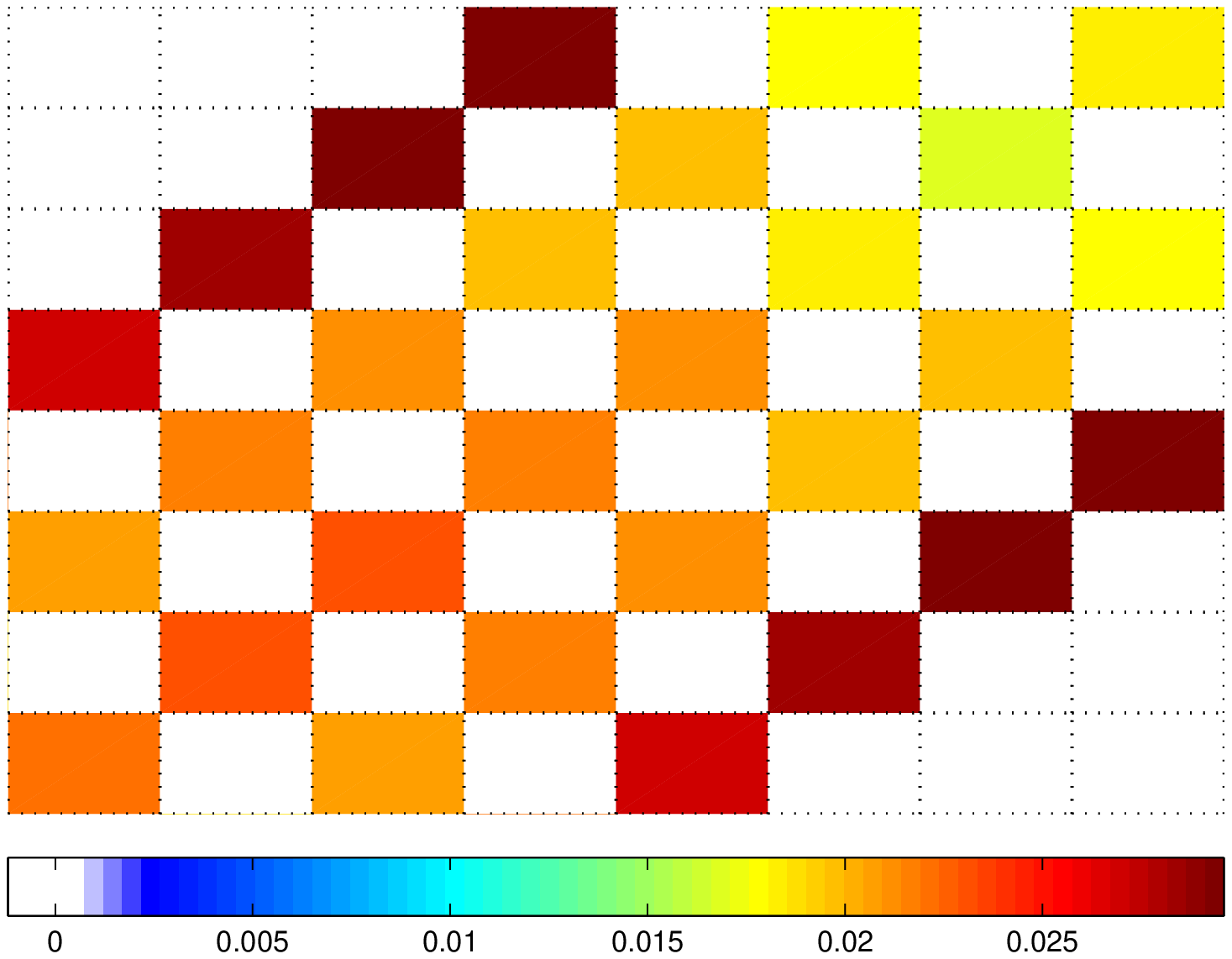}}}
\mbox{\resizebox{0.49\textwidth}{!}{\includegraphics[angle=0]{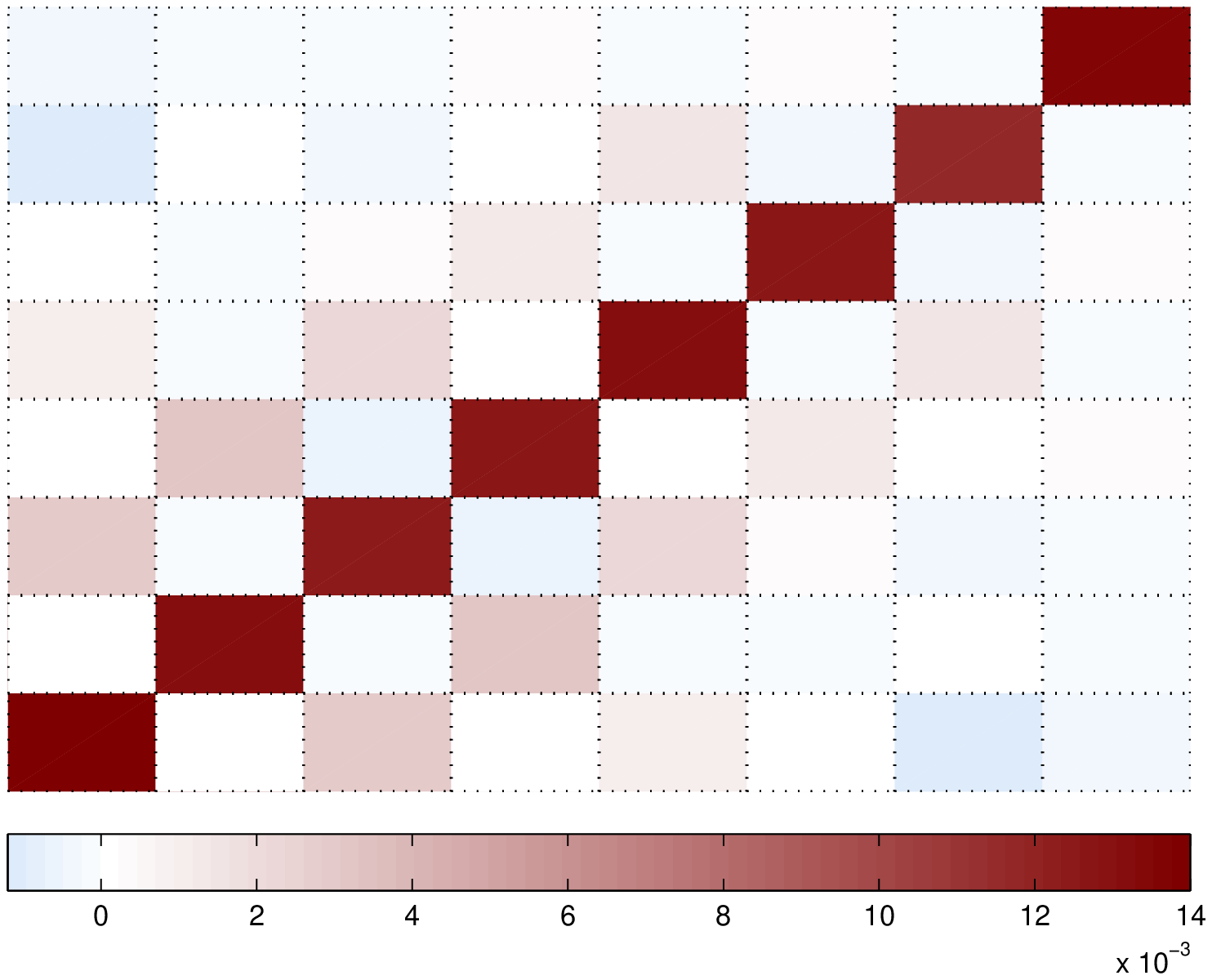}}}
\caption{\label{fig:cov_cl}  (Left) Covariance between $C_{\ell}$  for the ISW component of the power spectrum, using an AEDE model with $w=-0.2$, ${\bar\mu}_{\rm T}={\bar\mu}^{\rm min}$ and ${\bar\mu}_{\rm L}={\bar\mu}^{\rm min}+0.1$. (Right) Same quantity for the ACW model with $g_{\star}=1$.}
\end{figure}

Off-diagonal terms of the covariance $(C_{\ell_1}, C_{\ell_2})$ are also non-zero in anisotropic models. We calculate this quantity numerically by simulating a large number of realizations ($\sim 10^{6}$) of the sample covariance matrix. The result of these simulations for the AEDE and ACW models are shown in Fig.~\ref{fig:cov_cl}. For visual clarity, we plot the covariance relative to the isotropic value by
\beq
\left( {\rm Cov} (C_{\ell_1}, C_{\ell_2})- \delta_{\ell_1 \ell_2} \, \sigma^{2 \,\, \rm ISO}_{C_{\ell_1}} \right)/ \delta_{\ell_1 \ell_2} \, \sigma^{2 \,\, \rm ISO}_{C_{\ell_1}}\,.
\eeq
One can see that, relative the the main diagonal, the covariance $(C_{\ell_1}, C_{\ell_2})$ is larger for the AEDE model. This is apparent by examining the size of the $\ell_1-\ell_2$ correlations in Fig.~\ref{fig:cov_modes}.

%Simulations show the variance does not depend on the couplings between different $\ell$ modes, only the couplings between $m$ modes for each $\ell$.

We now turn our attention to the question of whether the anisotropy is detectable by any observation. The optimal probability of distinguishing between two models with covariance matrices $C^{A}$ and $C^{B}$, assuming model $A$ is correct, in a cosmic variance limited experiment is given by 
\beq \label{eqn:discrim}
\left \langle \ln \left( \frac{P(\{a_{\ell m} \} | A)}{P(\{a_{\ell m} \} | B)}\right) \right \rangle_{A} = -\frac{1}{2} \ \left[  {\rm tr} \left({\rm I}- \frac{C_{}^{(A)}}{C_{}^{(B)}} \right) + \ln \left( \frac{{\rm det} \, C_{}^{(A)}}{{\rm det} \, C_{}^{(B)}} \right) \right]\,.
\eeq
This is equivalent to the expression used in ref.~\cite{Battye:1999eq}, but generalized to the anisotropic case; the derivation of this equation is given in appendix~\ref{app:discrim}. In particular, we assume that the anisotropic model is correct, since we want to find the probability that it could be distinguished from the isotropic case. We also assume that both the isotropic and anisotropic model have the same $C_{\ell}$, defined by eqn.~(\ref{eqn:cl}), since we wish to estimate the significance of only the anisotropic terms, not the overall  power at each $\ell$. In this case, the first term on the right hand side is of eqn.~(\ref{eqn:discrim}) is zero, so the probability is simply the ratio of the determinant of the covariance matrices. 

\begin{figure}
\centering
\mbox{\resizebox{0.49\textwidth}{!}{\includegraphics[angle=0]{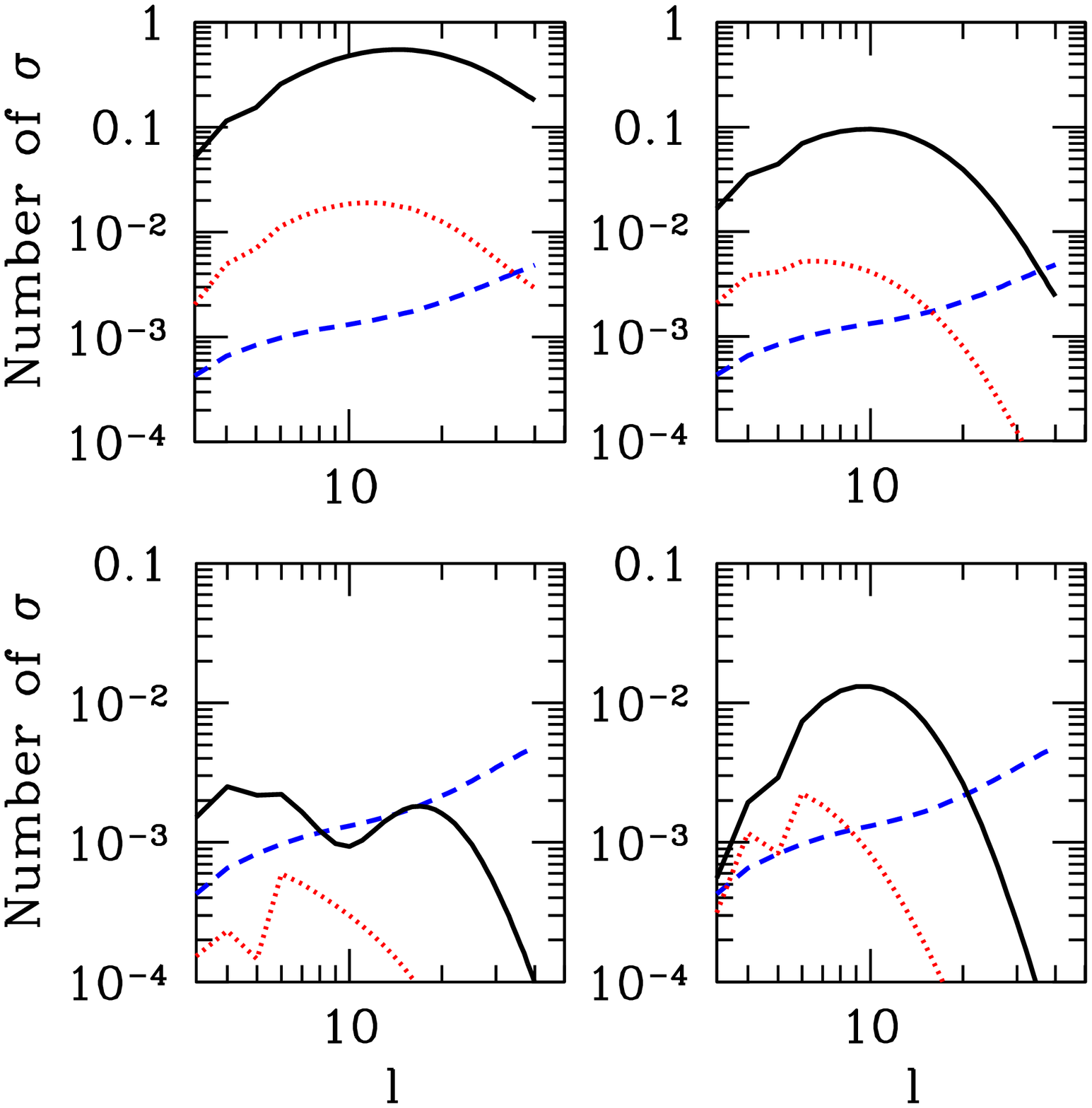}}}
\mbox{\resizebox{0.49\textwidth}{!}{\includegraphics[angle=0]{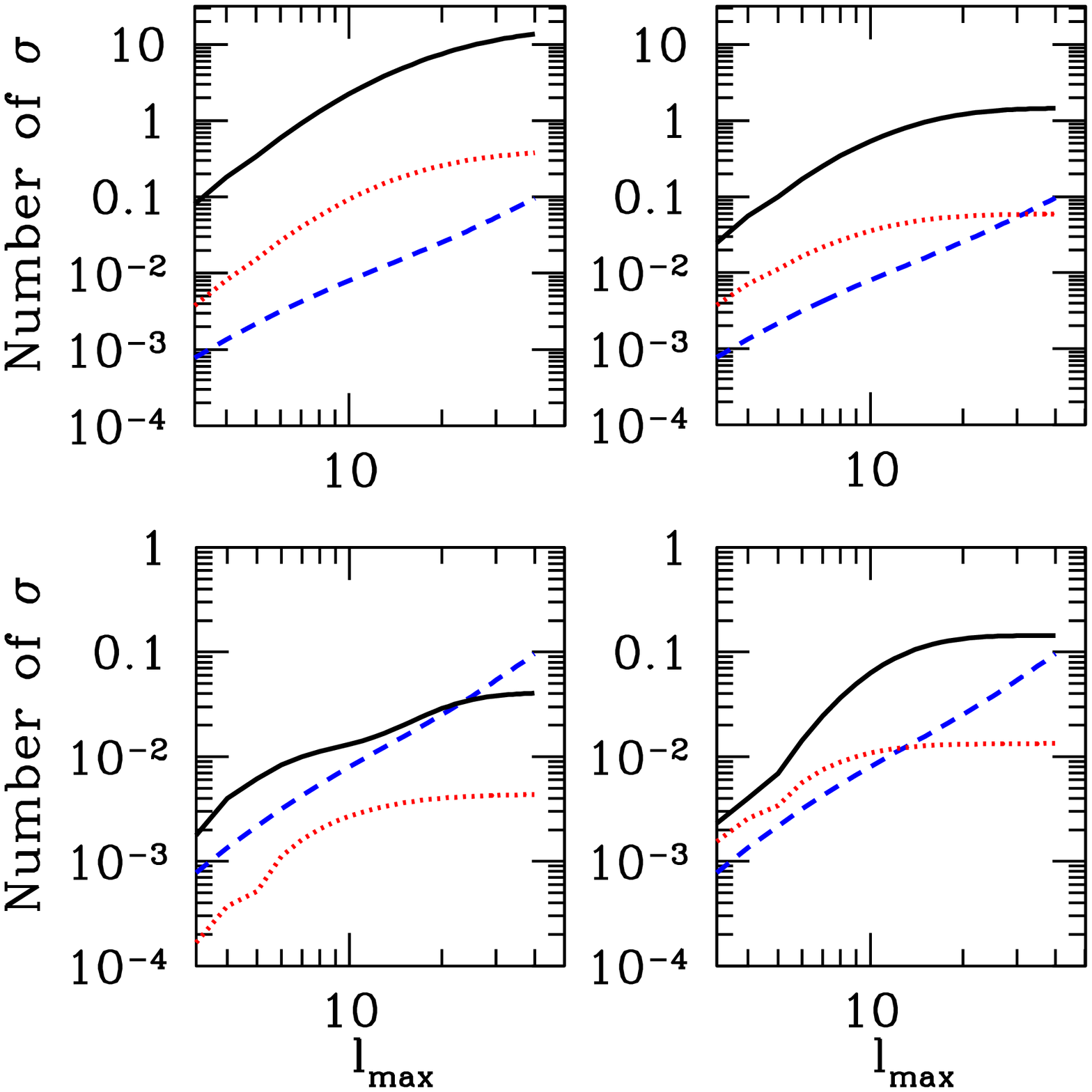}}}
\caption{\label{fig:prob} Probability of distinguishing between an anisotropic and isotropic model each with the same $C_{\ell}$. The labeling is the same as for Fig.~\ref{fig:tt}, where for the ACW model we use a value of $g_{\star}=0.1$. On the left, we show the contribution from each $\ell$ mode, and on the right the cumulative probability.}
\end{figure}

In Fig.~\ref{fig:prob} we show the both cumulative probability and contribution from each $\ell$ mode for AEDE and ACW models. The cumulative probability in the ACW model increases as roughly $\ell_{\rm max}^2$ as one would expect for an equal contribution from each mode, while the contribution from each $\ell$ decreases markedly for $\ell > 10 -20$ in the AEDE model. An extrapolation by eye suggests that a several $\sigma$ `detection' would be possible for the ACW model with an $\ell_{\rm max}$ of several hundred, which is compatible with the results of ref.~\cite{Groeneboom:2008fz}. The AEDE models, however, do not reach this threshold when considering the aniostropy created by the total ISW + LSS components.

Up to this point, we have only considered models with an scalar initial conditions. Certain models of inflation, however, are also expected to produce tensor fluctuations. The magnitude of these modes are model dependent, but power-law models of inflation, for example, have an observationally derived  2$\sigma$ limit on the tensor-to-scalar ratio of $r<0.42$, at a pivot scale $k_0=0.002 \, {\rm Mpc}^{-1}$~\cite{Dunkley:2008ie}. 

The effect of isotropic elastic dark energy on tensor fluctuations was discussed in ref.~\cite{Battye:2007aa}. The additional source of anisotropic stress damps the evolution of tensor modes (which decay through horizon crossing), causing a reduction in CMB power on large-scales. In the anisotropic case, this damping will be dependent on the direction of the wave vector, causing correlations in the covariance matrix. Moreover, in the same way as vector and tensor modes are excited in the scalar case, scalar and vectors modes will be generated from purely tensor initial conditions. 

The probability of detecting anisotropy in AEDE models will change if tensor modes are present, so we have investigated this possibility using the same values of $w$ as in the scalar case. Since the initial conditions for each of the scalar and tensor cases are uncorrelated, one can simply add the two covariance matrices.  We find, however, that the detection probability does not change significantly for for several reasons.  Firstly, the anisotropic contributions to the tensor covariance matrix along the main diagonal are actually {\em opposite} in sign to the scalar case -- the nature of this can be seen by computing the individual components of eqn.~(\ref{eqn:ang}). The tensor-tensor term, which is dominant in this case, distributes power along the main diagonal differently to the scalar-scalar term (for example, for the scalar-scalar term, more power goes to the $m=0$ modes). Secondly, the overall amplitude of the tensors are subdominant compared to scalars for observationally allowed values of $r$. Finally, we find that the probability of distinguishing between an anisotropic and isotropic model for the tensor covariance matrix also peaks around $\ell \sim 10$. In summary, this fall off in anisotropic power on smaller scales appears to be a large stumbling block for this class models.

%----------------- CONCLUSIONS -----------------------

\section{Conclusions}

We have developed a formalism for computing the CMB covariance matrix for models with an anisotropic dark energy component. It appears that the level of off-digonal terms which can be created by models where the pressure tensor is isotropic, but the perturbed energy-momentum tensor is not (via the elasticity tensor) is too small to be detected. This is true even in the most optimistic scenario where the values of $w$, $\mu_{\rm T}$ and $\mu_{\rm L}$ are chosen to give the largest possible signal. This is most explicitly illustrated by the fact that the dotted lines in Fig.~\ref{fig:prob} appear to top out a level below a 1-$\sigma$ detection when $\ell_{\rm max}>10$. Note the contrast with the ACW model for which the curves increase $\propto \ell_{\rm max}^2$ and therefore inclusion of a sufficiently large number of $\ell$-modes could lead to a detection as has been claimed~\cite{Groeneboom:2008fz}. This is because the anisotropic impact of the dark energy only comes from the ISW effect which is only strong on the largest scales, whereas in the ACW model it is present on all scales.

This negative result is something of a disappointment since this idea was one of the few which could naturally lead to a non-diagonal covariance matrix from scalar density fluctuations created during inflation. Clearly for such a model to have a change of being detected, one needs to generate larger off-diagonal terms in the covariance model. This may be possible in more general dark energy models of this kind which have an anisotropic pressure tensor as well as an anisotropic elasticity tensor. This will require the model to be embedded in a more general anistropic spacetime such as Bianchi I. This possibility is presently under consideration.

%----------------- ACKNOWLEDGMENTS -----------------------

\section*{Acknowledgments} This research was supported by the Natural Sciences and Engineering Research Council of Canada.  The calculations were performed on computing infrastructure purchased with funds from the Canadian Foundation for Innovation and the British Columbia Knowledge Development Fund. We thank Douglas Scott, Kris Sigurdson and Jim Zibin for useful conversations.

%----------------- APPENDIX -----------------------
\appendix

\section{Calculation of $I_{\ell m}^{(n)}$  \label{app:cov}}

The standard expansion of  a plane wave gives
\beq
e^{ix\mu}=\sum_{\ell}i^{\ell}(2\ell+1)j_\ell(x)P_{\ell}(\mu)=4\pi\sum_{\ell,m}i^{\ell}j_{\ell}(x)Y_{\ell m}(\hat{\bf k})Y_{\ell m}^*(\hat{\bf n})\,.
\eeq
This can be used in to show that 
\beq 
I_{\ell m}^{(0)}(x,\hat{\bf k})=4\pi i^{\ell}j_{\ell}(x)Y_{\ell m}({\hat{\bf k}})\,.
\eeq
We will use $(\theta_{\hat k},\phi_{\hat k})$ will refer to polar angles in $k$-space. Using these angles we can write the basis vectors~(\ref{eqn:basis}) as
\begin{eqnarray}
\hat{k}_i&=&(\sin\theta_{\hat k}\cos\phi_{\hat k},\sin\theta_{\hat k}\sin\phi_{\hat k},\cos\theta_{\hat k})\,,\nonumber \\ 
e^2_i&=&(\sin\phi_{\hat k},-\cos\phi_{\hat k},0)\,, \nonumber \\
e^3_i&=&(\cos\theta_{\hat k}\cos\phi_{\hat k},\cos\theta_{\hat k}\sin\phi_{\hat k},-\sin\theta_{\hat k})\,.
\end{eqnarray}
From these definitions, we can deduce that 
\begin{subequations}
\begin{eqnarray}
{\partial\hat{ k}_i \over \partial\theta_{\hat k}}&=& e_3^i \,,\qquad {\partial\hat{k}_i \over \partial\phi_{\hat k}}=-e^2_i  \sin\theta_{\hat k} \,,\\
{\partial e^2_i \over \partial\theta_{\hat k}}&=&0\,,\qquad {\partial e^2_i \over \partial\phi_{\hat k}}= \hat{k}_i \sin\theta_{\hat k} + e^3_i \cos\theta_{\hat k} \,,\\
{\partial e^3_i \over \partial\theta_{\hat k}}&=&-\hat{k}_i \,,\qquad {\partial e^3_i \over \partial\phi_{\hat k}}=- e^2_i \cos\theta_{\hat k} \,.
\end{eqnarray}
\end{subequations}
By computing derivatives with respect to $\theta$ and $\phi$, one can easily show that 
\begin{subequations}
\begin{eqnarray}
I^{(-1)}_{\ell m}&=&- {1\over ix\sin\theta_{\hat k}}{\partial\over\partial\phi_{\hat k}}I_{\ell m}^{(0)}=4\pi i^{\ell-1}{j_\ell(x)\over x}{1\over\sin\theta_{\hat k}}{\partial\over\partial\phi_{\hat k}}Y_{\ell m}(\hat{\bf k})\,,\\
I^{(+1)}_{\ell m}&=&{1\over ix}{\partial\over\partial\theta_{\hat k}}I_{\ell m}^{(0)}=4\pi i^{\ell-1}{j_\ell(x)\over x}{\partial\over\partial\theta_{\hat k}}Y_{\ell m}(\hat{\bf k})\,,\\
I^{(-2)}_{\ell m}&=&{1\over ix}\left( {\partial\over\partial\theta_{\hat k}}I_{\ell m}^{(-1)}-{1\over\sin\theta_{\hat k}}{\partial\over\partial\phi_{\hat k}}I_{\ell m}^{(+1)}-\cot\theta_{\hat k} I_{\ell m}^{(-1)}\right)\\
&=&4\pi i^{\ell-2}{j_{\ell}(x)\over x^2}\left({\partial^2\over\partial\theta_{\hat k}^2}-\cot\theta_{\hat k}{\partial\over\partial\theta_{\hat k}}-{1\over\sin^2\theta_{\hat k}}{\partial^2\over \partial\phi^2_{\hat k}}\right)Y_{\ell m}(\hat{\bf k}) \,, \\
I^{(+2)}_{\ell m}&=&{2\over ix}{\partial\over\partial\theta_{\hat k}}I_{\ell m}^{(+1)}=8\pi i^{\ell -2}{j_{\ell}(x)\over x^2}{\partial\over\partial\theta_{\hat k}}\left({1\over\sin\theta_{\hat k}}{\partial\over\partial\phi_{\hat k}}\right)Y_{\ell m}(\hat{\bf k})\,. 
\end{eqnarray}
\end{subequations}
Using these expressions we can deduce that 
\beq 
I_{\ell m}^{(n)}=4\pi i^{\ell-|n|}{j_{\ell}(x)\over x^{|n|}} J_{\ell m}^{(n)}(\hat{\bf k})\,,
\eeq
where the functions $J_{\ell m}^{(n)}$ are defined by 
\begin{subequations}
\begin{eqnarray}
J_{\ell m}^{(0)}(\hat{\bf k})&=&Y_{\ell m}(\hat{\bf {k}})\,,\\
J_{\ell m}^{(-1)}(\hat{\bf k})&=& {1\over \sin\theta_{\hat k}}{\partial\over\partial\phi_{\hat k}}Y_{\ell m}({\hat{\bf k}})\,,\\
 J_{\ell m}^{(+1)}(\hat{\bf k})&=& {\partial\over\partial\theta_{\hat k}}Y_{\ell m}({\hat {\bf{k}}})\,,\\
J_{\ell m}^{(-2)}(\hat{\bf k})&=& \left({\partial^2\over\partial\theta_{\hat k}^2}-\cot\theta_{\hat k}{\partial\over\partial\theta_{\hat k}}-{1\over\sin^2\theta_{\hat k}}{\partial^2\over\partial\phi_{\hat k}^2}\right)Y_{\ell m}({\hat{\bf k}})\,, \\
J_{\ell m}^{(+2)}(\hat{\bf k})&=& 2{\partial\over\partial\theta_{\hat k}}\left({1\over \sin\theta_{\hat k}}{\partial\over\partial\phi_{\hat k}}\right)Y_{\ell m}({\hat {\bf k}})\,.
\end{eqnarray}
\end{subequations}
These can be written in terms of the spin-s harmonics 
\begin{subequations}
\begin{eqnarray}
J^{(-1)}_{\ell m}(\hat{\bf k})&=&-\left[{(\ell+1)!\over (\ell-1)!}\right]^{1/2}{ _{+1}Y_{\ell m}({\hat{\bf k}})+ _{-1}Y_{\ell m}({\hat{\bf k}})\over 2 i}\,, \\
J^{(+1)}_{\ell m}(\hat{\bf k})&=&-\left[{(\ell+1)!\over (\ell-1)!}\right]^{1/2}{ _{+1}Y_{\ell m}({\hat{\bf k}})- _{-1}Y_{\ell m}({\hat{\bf k}})\over 2 }\,, \\
J^{(-2)}_{\ell m}(\hat{\bf k})&=&\left[{(\ell+2)!\over (\ell-2)!}\right]^{1/2}{ _{+2}Y_{\ell m}({\hat{\bf k}})+ _{-2}Y_{\ell m}({\hat{\bf k}})\over 2}\,, \\
J^{(_+2)}_{\ell m}(\hat{\bf k})&=&\left[{(\ell+2)!\over (\ell-2)!}\right]^{1/2}{ _{+2}Y_{\ell m}({\hat{\bf k}})- _{-2}Y_{\ell m}({\hat{\bf k}})\over 2i}\,.
\end{eqnarray}
\end{subequations}
The spin-s harmonics can also be generated from the  series expansion
\begin{equation}
_s Y_{\ell m} = \left[ \frac{2\ell +1 }{4 \pi} \frac{(\ell+m)!}{(\ell+s)!} \frac{(\ell-m)!}{(\ell-s)!}\right]^{1/2} e^{i m \phi} \sin^{2\ell}(\theta/2) \sum_{r}^{\ell-s} (\ell-s, r) (\ell+s,r+s-m) (-1)^{\ell-r-s} \cot^{2r+s-m}(\theta/2)\,,
\end{equation}
where $(X,Y)$ is the binomial coefficient. 

\section{Calculation of discrimination equation}  \label{app:discrim}

Given two models with covariance matrices $C^{(A)}$ and $C^{(B)}$, and the data vector ${\bf a}$, the ratio of probabilities between models is
\beq
\ln \left( \frac{P(\{a\} | A)}{P(\{a\} | B)}\right) = -\frac{1}{2} \ \left[ {\bf a}^{\rm T \star} C^{(A) \, -1} {\bf a} - {\bf a}^{\rm T \star} C^{(B) \, -1} {\bf a} + \ln \left( \frac{{\rm det} \, C_{}^{(A)}}{{\rm det} \, C_{}^{(B)}} \right) \right]\,.
\eeq
We assume that $C^{(A)}$ is the true covariance matrix so that ${\bf a} = A {\bf \zeta}$, where $A (A^{\rm T})^{\star}=C^{A}$ is the Cholesky decomposition of the covariance matrix . For Gaussian initial conditions one has that $\langle \zeta_{t}^{\star} \zeta_{t^{\prime}} \rangle= \delta_{t t^{\prime}}$, so that
\begin{subequations}
\begin{eqnarray}
\left \langle \ln \left( \frac{P(\{a\} | A)}{P(\{a\} | B)}\right) \right \rangle_{A} &=& -\frac{1}{2} \ \left[ {\rm tr} \left( A^{\rm T \star} C^{(A) \, -1} A \right) - {\rm tr} \left( A^{\rm T \star} C^{(B) \, -1} A \right)  + \ln \left( \frac{{\rm det} \, C_{}^{(A)}}{{\rm det} \, C_{}^{(B)}} \right) \right]\,, \\ 
&=& -\frac{1}{2} \ \left[  {\rm tr} \left({\rm I}- \frac{C_{}^{(A)}}{C_{}^{(B)}} \right) + \ln \left( \frac{{\rm det} \, C_{}^{(A)}}{{\rm det} \, C_{}^{(B)}} \right) \right]\,.
\end{eqnarray}
\end{subequations}


\begin{thebibliography}{99}
\newcommand{\prlet}{Phys.\ Rev.\ Lett.}
\newcommand{\npb}{Nucl.\ Phys.\ B}
\newcommand{\pletb}{Phys.\ Lett.\ B}
\newcommand{\prevd}{Phys.\ Rev.\ D}
\newcommand{\jhep}{J.\ High Energy Phys.}
\newcommand{\cqg}{Class.\ Quant.\ Grav.}
\newcommand{\jast}{Astrophys. \ J.}

\bibitem{Hinshaw:2008kr}
G. Hinshaw {\em et al}, {\em Astrophys. \ J.\.Supp.} {\bf 180}, (2009) 225 [arXiv:0803.0732].

\bibitem{Yadav:2007yy}
A. P. S. Yadav and B. D. Wandelt, {\em \prlet} {\bf 100}, (2008) 181301 [arXiv:0712.1148].

\bibitem{Smoot:1992td}
G. F. Smoot {\em et al}, {\em \jast} {\bf 396}, (1992) L1.

\bibitem{Hinshaw:2006ia}
G. Hinshaw {\em et al}, {\em Astrophys. \ J.\.Supp.} {\bf 170}, (2007) 288 [astro-ph/0603451].

\bibitem{Spergel:2003cb}
D. Spergel {\em et al}, {\em Astrophys. \ J.\.Supp.} {\bf 148}, (2003) 175 [astro-ph/0302209].

\bibitem{Copi:2006tu}
C. J. Copi, D. Huterer, D. Schwarz and G. D. Starkman, {\em \prevd} {\bf 75}, (2007) 023507 [astro-ph/0605135].

\bibitem{Eriksen:2003db}
H. K. Eriksen, F. K. Hansen, A. J. Banday, K. M. Gorski and P. B. Lilje, {\em \jast} {\bf 605}, (2004) 14 [astro-ph/0307507].

\bibitem{Hansen:2004vq}
F. K. Hansen, A. J. Banday and K. M. Gorski, (2004) [astro-ph/0404206].

\bibitem{Jaffe:2005pw}
T. R. Jaffe, A. J. Banday, H. K. Eriksen, K. M. Gorski and F. K. Hansen, {\em \jast} {\bf 629}, (2005) L1 [astro-ph/0503213].

\bibitem{Vielva:2003et}
P. Vielva, E. Martinez-Gonzalez, R. Barreiro, J. L. Sanz and L. Cayon, {\em \jast} {\bf 609}, (2004) 22 [astro-ph/0310273].

\bibitem{Cruz:2004ce}
M. Cruz, E. Martinez-Gonzalez, P. Vielva and L. Cayon, {\em MNRAS} {\bf 356}, (2005) 29 [astro-ph/0405341].

\bibitem{de OliveiraCosta:2003pu}
A. de Oliveira-Costa, M. Tegmark, M. Zaldarriaga and A. Hamilton, {\em \prevd} {\bf 69}, (2004) 063516 [astro-ph/0307282].

\bibitem{Copi:2003kt}
C. J. Copi, D. Huterer and G. D. Starkman, {\em \prevd} {\bf 70}, (2004) 043515 [astro-ph/0310511].

\bibitem{Schwarz:2004gk}
D. J. Schwarz, G. D. Starkman, D. Huterer and C. J. Copi, {\em \prlet} {\bf 93}, (2004) 221301 [astro-ph/0403353].

\bibitem{Prunet:2004zy}
S. Prunet, J. P. Uzan, F. Bernardeau and T. Brunier, {\em \prevd} {\bf 71}, (2005) 083508 [astro-ph/0406364].

\bibitem{Land:2005ad}
K. Land and J. Magueijo, {\em \prlet} {\bf 95}, (2005) 071301 [astro-ph/0502237].

\bibitem{Land:2006bn}
K. Land and J. Magueijo, {\em MNRAS} {\bf 378}, (2007) 153 [astro-ph/0611518].

\bibitem{Buniy:2005qm}
R. V. Buniy, A. Berera and T. W. Kephart, {\em \prevd} {\bf 73}, (2006) 063529 [hep-ph/051115].

\bibitem{Gumrukcuoglu:2006xj}
A. E. Gumrukcuoglu, C. R. Contaldi and M. Peloso,  (2006) [astro-ph/0608405].

\bibitem{Ackerman:2007nb}
L. Ackerman, S. M. Carroll and M. B. Wise, {\em \prevd} {\bf 75}, (2007) 083502 [astro-ph/0701357].

\bibitem{Seljak:1996is}
U. Seljak and M. Zaldarriaga, {\em \jast} {\bf 469}, (1996) 437 [astro-ph/9603033].

\bibitem{Lewis:1999bs}
A. Lewis, A. Challinor and A. Lasenby, {\em \jast} {\bf 538}, (2000) 473 [astro-ph/9911177].

\bibitem{Groeneboom:2008fz}
N. E. Groeneboom and H. K. Eriksen, {\em \jast} {\bf 690}, (2009) 1807 [arXiv:0807.2242].

\bibitem{Himmetoglu:2008hx}
B. Himmetoglu, C. R. Contaldi and M. Peloso, (2008) [arXiv:0812.1231].

\bibitem{Copeland:2006wr}
E. J. Copeland, M. Sami and S. Tsujikawa, hep-th/0603057.

\bibitem{Hu:1998kj}
W. Hu,  {\em \jast} {\bf 506}, (1998) 485 [astro-ph/9801234].

\bibitem{Koivisto:2005mm}
T. Koivisto and D. F. Mota,  {\em \prevd} {\bf 73}, (2006) 083502 [astro-ph/0512135].

\bibitem{Mota:2007sz}
D. F. Mota, J. R. Kristiansen, T. Koivisto and N. E. Groeneboom, {\em MNRAS} {\bf 382}, (2007) 793 [arXiv:0708.0830].

\bibitem{Koivisto:2007bp}
T. Koivisto and D. F. Mota, {\em \jast} {\bf 679}, (2008) 1 [arXiv:0707.0279].

\bibitem{Koivisto:2008ig}
T. Koivisto and D. F. Mota, {\em JCAP} {\bf 0806}, (2008) 18 [arXiv:0801.3676].

\bibitem{Battye:2006mb}
R. A. Battye and A. Moss, {\em \prevd} {\bf 74}, (2006) 041301 [astro-ph/0602377].

\bibitem{Carter:1972cq}
B. Carter and H. Quintana, {\em Proc. Roy. Soc.} {\bf A331}, (1972) 57.

\bibitem{Carter:1977qf}
B. Carter and H. Quintana, {\em \prevd} {\bf 16}, (1977) 2928.

\bibitem{Carter:1980c}
B. Carter, {\em Proc. Roy. Soc.} {\bf A372}, (1980) 169.

\bibitem{Carter:1982xm}
B. Carter, gr-qc/0102113.

\bibitem{Battye:2007aa}
R. A. Battye and A. Moss, {\em \prevd} {\bf 76}, (2007) 023005 [astro-ph/0703744].

\bibitem{Bucher:1998mh}
M. Bucher and D.N. Spergel, {\em \prevd} {\bf 60}, (1999) 043505 [astro-ph/9812022].

\bibitem{Battye:1999eq}
R.A. Battye, M. Bucher and D.N. Spergel, astro-ph/9908047.

\bibitem{landau:1959} 
L.D. Landau, E.M. Lifshitz, {\it Theory of Elasticity} (Pergamon, London, 1959). 

\bibitem{Niarchou:2007nn}
A. Niarchou and A. Jaffe, {\em \prlet} {\bf 99}, (2007) 081302 [astro-ph/0702436].

\bibitem{Battye:2005ik}
R. A. Battye, E. Chachoua and A. Moss, {\em \prevd} {\bf 73}, (2006) 123528 [hep-th/0512207].

\bibitem{Lifshitz:1963ps}
E.~M. Lifshitz and I.~M. Khalatnikov, {\em Adv. Phys.} {\bf 12}, (1963) 185.

\bibitem{Hu:1997mn}
W. Hu, U. Seljak, M. White and M. Zaldarriaga,  {\em \prevd} {\bf 57}, (1998) 3290 [astro-ph/9709066].

\bibitem{Pereira:2007yy}
T. S. Pereira, C. Pitrou and J. P. Uzan, {\em JCAP} {\bf 0709}, (2007) 006 [arXiv:0707.0736].

\bibitem{Dunkley:2008ie}   
J. Dunkley {\em et al}, {\em Astrophys. \ J.\.Supp.} {\bf 180}, (2009) 306 [arXiv:0803.0586].

\bibitem{Weller:2003hw} 
J. Weller and A. Lewis, {\em MNRAS} {\bf 346} (2003) 987 [astro-ph/0307104].

\bibitem{Bean:2003fb} 
R. Bean and O. Dore, {\em \prevd} {\bf 69} (2004) 083503 [astro-ph/0307100]. 

\bibitem{Kowalski:2008ez}
M. Kowalski {\em et al}, {\em \jast} {\bf 686}, (2008) 749 [arXiv:0804.4142].

\bibitem{Ferreira:1997wd}
P. G. Ferreira and J. Magueijo, {\em \prevd} {\bf 56}, (1997)  4578 [astro-ph/9704052].



\end{thebibliography}
\end{document}